\documentclass[fleqn,usenatbib]{mnras}

\usepackage{newtxtext,newtxmath}

\usepackage[T1]{fontenc}

\DeclareRobustCommand{\VAN}[3]{#2}
\let\VANthebibliography\thebibliography
\def\thebibliography{\DeclareRobustCommand{\VAN}[3]{##3}\VANthebibliography}

\usepackage{graphicx}	
\usepackage{amsmath}	

\newcommand{\Ion}[2]{#1\,\textsc{#2}}
\newcommand{\kms}{\mbox{km\,s$^{-1}$}}

\title[Age-Activity-Rotation relation]{Main sequence companions to white dwarfs II: the age-activity-rotation relation from a sample of \emph{Gaia} common proper motion pairs}

\author[A. Rebassa-Mansergas et al.]{A. Rebassa-Mansergas$^{1,2}$\thanks{E-mail: alberto.rebassa@upc.edu}, J. Maldonado$^{3}$, R. Raddi$^{1}$, S. Torres$^{1,2}$, M. Hoskin$^{4}$, T. Cunningham$^{4}$, \newauthor  M.~A.~Hollands$^{5}$, J. Ren$^{6}$, B. T. G\"ansicke$^{4}$, P.-E. Tremblay$^{4}$, M. Camisassa$^{1}$\\
$^{1}$ Departament de F\'isica, Universitat Polit\`ecnica de Catalunya, c/Esteve Terrades 5, 08860 Castelldefels, Spain\\
$^{2}$ Institute for Space Studies of Catalonia, c/Gran Capit\`a 2--4, Edif. Nexus 104, 08034 Barcelona, Spain\\
$^{3}$ INAF - Osservatorio Astronomico di Palermo, Piazza del Parlamento 1, 90134 Palermo, Italy\\
$^{4}$ Department of Physics, University of Warwick, Coventry, CV4 7AL, UK\\
$^{5}$ Department of Physics and Astronomy, University of Sheffield, Sheffield, S3 7RH, UK\\
$^{6}$ CAS Key Laboratory of Space Astronomy and Technology, National Astronomical Observatories, Chinese Academy of Sciences, Beijing 100101, China\\
}

\date{Accepted XXX. Received YYY; in original form ZZZ}

\pubyear{2021}

\begin{document}
\label{firstpage}
\pagerange{\pageref{firstpage}--\pageref{lastpage}}
\maketitle

\begin{abstract}
Magnetic activity and rotation are related to the age of low-mass main
sequence  stars.  To  further constrain  these relations,  we study  a
sample  of 574  main sequence  stars members  of common  proper motion
pairs with white dwarfs,  identified thanks to \emph{Gaia} astrometry.
We use the white dwarfs as  age indicators, while the activity indexes
and  rotational  velocities  are   obtained  from  the  main  sequence
companions using standard  procedures.  We find that  stars older than
5\,Gyr do not display  H$\alpha$ nor \Ion{Ca}{ii}~H\&K emission unless
they  are fast  rotators due  to tidal  locking from  the presence  of
unseen companions and that the  rotational velocities tend to decrease
over time, thus supporting  the so-called gyrochronology.  However, we
also  find moderately  old  stars ($\simeq$2-6\,Gyr)  that are  active
presumably because they rotate faster than they should for their given
ages.   This  indicates  that  they may  be  suffering  from  weakened
magnetic braking or that they  possibly evolved through wind accretion
processes in the past.  The activity fractions that we measure for all
stars  younger than  5\,Gyr  range  between $\simeq$10--40  per\,cent.
This is line with the expectations, since our sample is composed of F,
G, K  and early M stars,  which are thought to  have short ($<2$\,Gyr)
activity lifetimes.  Finally, we observe that the H$\alpha$ fractional
luminosities  and  the  R'$_\mathrm{HK}$  indexes for  our  sample  of
(slowly        rotating)       stars        show       a        spread
($-4>$log$(L_\mathrm{H\alpha}/L_\mathrm{bol}$);  log(R'$_\mathrm{HK}$)
> -5) typically found in inactive M stars or weakly active/inactive F,
G, K stars.
\end{abstract}

\begin{keywords}
(stars:) binaries: general -- stars: low-mass -- stars: activity --
  (stars:) white dwarfs
\end{keywords}



\section{Introduction}

Magnetic  field  generation  and  angular momentum  evolution  are  of
fundamental  importance  to  understand  the  physical  properties  of
low-mass main sequence  stars. These processes have  been studied over
several decades and the current picture can be summarised as follows.

In partially  convective stars  of spectral  types late  F, G,  K, and
early M,  magnetic fields arise  due to a combination  of differential
rotation and convection, the  so-called $\Omega$ and $\alpha$ effects,
respectively   \citep{parker55-1,    leighton69-1,   spiegel+zahn92-1,
  charbonneau05-1,  browning08-1}.  This   $\alpha-\Omega$  dynamo  is
thought  to  generate  the  magnetic fields  at  the  tachocline  (the
transition region between the radiative and convective zones) of these
stars.  On  the other hand,  fully convective stars of  spectral types
later than  $\simeq$M3 do  not have  a tachocline  and are  thought to
rotate  as rigid  bodies \citep{barnesetal05-1},  hence an  $\alpha^2$
dynamo  is expected  to generate  the magnetic  fields in  these stars
\citep{raedleretal90-1,  chabrier+kueker06-1}.  A  $\gamma^2$  dynamo,
based only  on the turbulent  cross-helicity effect, can  also explain
the  generation   of  magnetic   fields  in  fully   convective  stars
\citep{Pipin2018}.

\begin{table*}
   \centering
  \caption{In this table  we provide the number of  WDMS binaries that
    are part of the full sample,  and that are part of subsamples with
    available  white dwarf  total ages,  with available  main sequence
    rotational  velocities,  with  available main  sequence  H$\alpha$
    fractional  luminosities   and  with   available  R'$_\mathrm{HK}$
    indexes.   For  each subsample  we  also  indicate the  number  of
    objects that pass the cuts we implemented (age uncertainties under
    1\,Gyr,  $v_\mathrm{rot}\sin  i$ errors  below  1  km/s and  $\log
    (L_\mathrm{H\alpha}/L_\mathrm{bol})$   uncertainties  lower   than
    0.75) and of  those, which ones are formed by  M dwarfs (effective
    temperatures lower or equal than  3800\,K and associated to errors
    lower than 120\,K)  or by FGK stars  (effective temperature higher
    than  3800\,K and  associated to  errors lower  than 120\,K).  The
    coordinates and  parameters derived in  this work for  each object
    are given in a supplementary material table.}
  \label{t-summary}
\begin{tabular}{cccccc} 
  \hline
& Full sample & Sample with ages & Sample with  $v_\mathrm{rot}\sin i$ & Sample with $L_\mathrm{H\alpha}/L_\mathrm{bol}$ & Sample with R'$_\mathrm{HK}$ index\\
 \hline
Total number    &   574   &   431  &  195  &  527 & 111\\
Pass cut        &   515   &   201  &  102  &  340 & 111\\
Are M dwarfs    &   201   &    76  &   0   &  127 &  19\\
Are FGK stars   &   314   &   125  &  102  &  213 &  92\\
\hline
  \end{tabular}                 
  \end{table*}

Observationally, it has been found that for partially convective stars
the efficiency of  magnetic field generation, which  results in higher
levels of  magnetic activity, is  strongly correlated with  the Rossby
number,   defined   as    $R_{0}   =   P_\mathrm{rot}/\tau_0$,   where
$P_\mathrm{rot}$  is  the  rotational   period  and  $\tau_0$  is  the
convective       overturn        timescale       \citep{noyesetal84-1,
  Garrafo2018}. Thus, for  a fixed $\tau_0$, stars  that rotate faster
display   more  signs   of   magnetic  activity   at  their   surfaces
\citep{wilson66-1,  kraft67-1,  hartmann+noyes87-1}.  This  connection
between rotation and  the efficiency of generating  magnetic fields is
also  expected for  fully convective  stars \citep{durney+stenflo72-1,
  chabrier+kueker06-1,   Shulyak2017,   Wright2018},   although   some
observational  studies  point  out  that  this does  not  have  to  be
necessarily  the case  \citep[e.g.][]{west+basri09-1}. In  particular,
\citet{mohantyetal02-1} and  \citet{reiners+basri10-1} found  no clear
correlation between  rotation and  activity for  ultracool M  stars of
spectral  types M7--9.5  and  \citet{Kiman2021}  suggest a  decreasing
activity trend for  M7 stars. It has to be  also emphasised that, even
though a  faster rotation generally  implies higher signs  of magnetic
activity,  both the  strength of  the  magnetic fields  and the  X-ray
emission   of    the   stars   saturate   at    $R_{0}   \simeq   0.1$
\citep{pizzolatoetal03-1,        reinersetal09-1,       Astudillo2017,
  Magaudda2020, Pineda2021}.

Given that low-mass  main sequence stars suffer  from magnetic braking
in which  angular momentum is  extracted from the  convective envelope
and lost through a magnetized  wind, their rotational periods decrease
in    time   \citep{skumanich72-1,    mestel84-1,   mestel+spruit87-1,
  kawaler88-1,    sillsetal00-1,   westetal08-1,    Meibom2015}.   The
timescales at which the rotational periods decrease have been observed
to significantly change  depending on the mass of the  star. Thus, the
rotational  braking is  considerably faster  for partially  convective
stars  than  for   fully  convective  stars  \citep{reiners+basri08-1,
  browningetal10-1,   Schreiber2010,  Zorotovic2016}   and  lower-mass
partially  convective stars  spin down  slower than  higher-mass stars
\citep{barnes03-1, barnes+kim10-1}. However, it  is not clear yet what
physical mechanisms are behind this disruption of magnetic braking for
fully convective stars. Some studies  suggest a change in the magnetic
field       topology       \citep{donatietal08-1,       morinetal08-1,
  reiners+basri09-1, morinetal10-1,  Reville2015} or a drop  in radius
\citep{reiners+mohanty12-1}  as possible  causes.  Due  to the  strong
correlation  between  rotation  and  activity, age  and  activity  are
correlated  too \citep{Houdebine2017},  and  it is  expected that  the
activity lifetimes  (i.e. the time  a star is magnetically  active) of
fully  convective  stars are  considerably  longer  than of  partially
convective  stars,  a  hypothesis  that   seems  to  be  confirmed  by
observations                              \citep[e.g.][]{westetal08-1,
  rebassa-mansergasetal13-1}. However,  it has  to be  emphasised that
measuring stellar ages,  and hence activity lifetimes,  is a difficult
endeavour subject  to substantial  uncertainties \citep{soderblom10-1,
  anguianoetal10-1}.
 
Dating low-mass main sequence stars to study the age-activity relation
relies (1) on  the analysis of the \Ion{Ca}{ii}~H\&K lines  as a proxy
of  age   \citep{Mamajek+Hillenbrand2008,  Pace2013},  (2)   on  their
kinematic    properties   \citep{Zhao2015,    Newton2016,   Angus2020,
  Yuxi2021}, (3) on asteroseismology \citep{Metcalfe2012, Chaplin2014,
  Silva2017, Booth2020},  (4) on  using the X-ray  emission as  an age
estimator \citep{Mamajek+Hillenbrand2008,  Maldonado2010}; (5)  on the
presence   of   lithium  in   the   spectra   to  date   young   stars
\citep{Gutierrez-Albarran2020};   and   (6)   on   isochrone   fitting
\citep{Angus2019}. Using  the rotation of  a given star to  derive its
age, the so-called gyrochronology, is  also a widely used technique to
estimate stellar ages  \citep{Epstein2014, Angus2015}.  However, there
exists  evidence  that  stars  that are  halfway  through  their  main
sequence lifetimes  suffer from  weakened magnetic  braking and,  as a
consequence,  are anomalously  rapidly rotating  \citep{vanSaders2016,
  Hall2021}.  Since only  stars that are younger than 1  Gyr have been
used  as calibrators  \citep{Barnes2010,  Garcia2014, Angus2015},  the
technique of gyrochronology should be used with caution for stars that
are more  evolved than  the Sun, unless  weakened magnetic  braking is
taken into account \citep{Sadeghi2017, Metcalfe2020}.

To provide additional observational inputs,  in this work we study the
age-rotation-activity  relations   using  binaries  consisting   of  a
low-mass  main  sequence  star  and  a  white  dwarf  (hereafter  WDMS
binaries). This  approach relies on  measuring accurate ages  from the
white dwarfs. To that end one  requires measuring the cooling ages and
masses of  the white dwarfs,  which together with  an initial-to-final
mass     relation     (e.g.      \citealt{Catalan08,     Cummings2018,
  Barrientos2021}),  allows determining  the main  sequence progenitor
masses  and their  lifetimes \citep{Fouesneau2019,  Qiu2020, Lam2020}.
This alternative  way of analysing the  age-rotation-activity relation
for low-mass  main sequence  stars has  been explored  in the  past by
\citet{morganetal12-1, Skinner2017} and by  our own group \citep[Paper
  I;][]{rebassa-mansergasetal13-1}   via   analysing   WDMS   binaries
identified     within     the     Sloan     Digital     Sky     Survey
\citep{rebassa-mansergasetal10-1,           rebassa-mansergasetal12-1,
  rebassa-mansergasetal13-2,               rebassa-mansergasetal16-1}.
Unfortunately, despite being the  most numerous and homogeneous sample
of  (spectroscopic) WDMS  binaries,  the SDSS  catalogue suffers  from
important  selection  effects \citep{rebassa-mansergasetal10-1}.   For
instance,  white  dwarfs  cooler than  $\simeq$10,000\,K  are  clearly
underrepresented.   This  effect  considerably reduces  the  range  of
studied white dwarf cooling times and biases the sample against larger
ages. Moreover, the  main sequence companions are  typically M dwarfs,
which  implies F,  G, and  K stars  cannot be  analysed and  therefore
assigned an  age. The  motivation of  the study  presented here  is to
surpass these  limitations by analysing  a sample of WDMS  binaries in
common  proper   motion  pairs,   identified  thanks   to  \emph{Gaia}
astrometry.   Since  the  two  stars are  widely  separated  they  are
expected to have evolved avoiding mass transfer episodes.  However, is
is  possible  that the  main  sequence  companions may  have  accreted
material  from the  white dwarf  progenitors via  stellar winds,  thus
gaining angular  momentum \citep{Jeffries1996, Boffin2015},  which may
alter the  expected age-rotation-activity  relations \citep{Zurlo2013,
  Leiner2018,  Bowler2021, Gratton2021}.   In  any case,  they can  be
independently  observed,  which  allows targeting  both  cooler  white
dwarfs and higher-mass main sequence stars.

\begin{figure}
    \centering
    \includegraphics[angle=-90,width=\columnwidth]{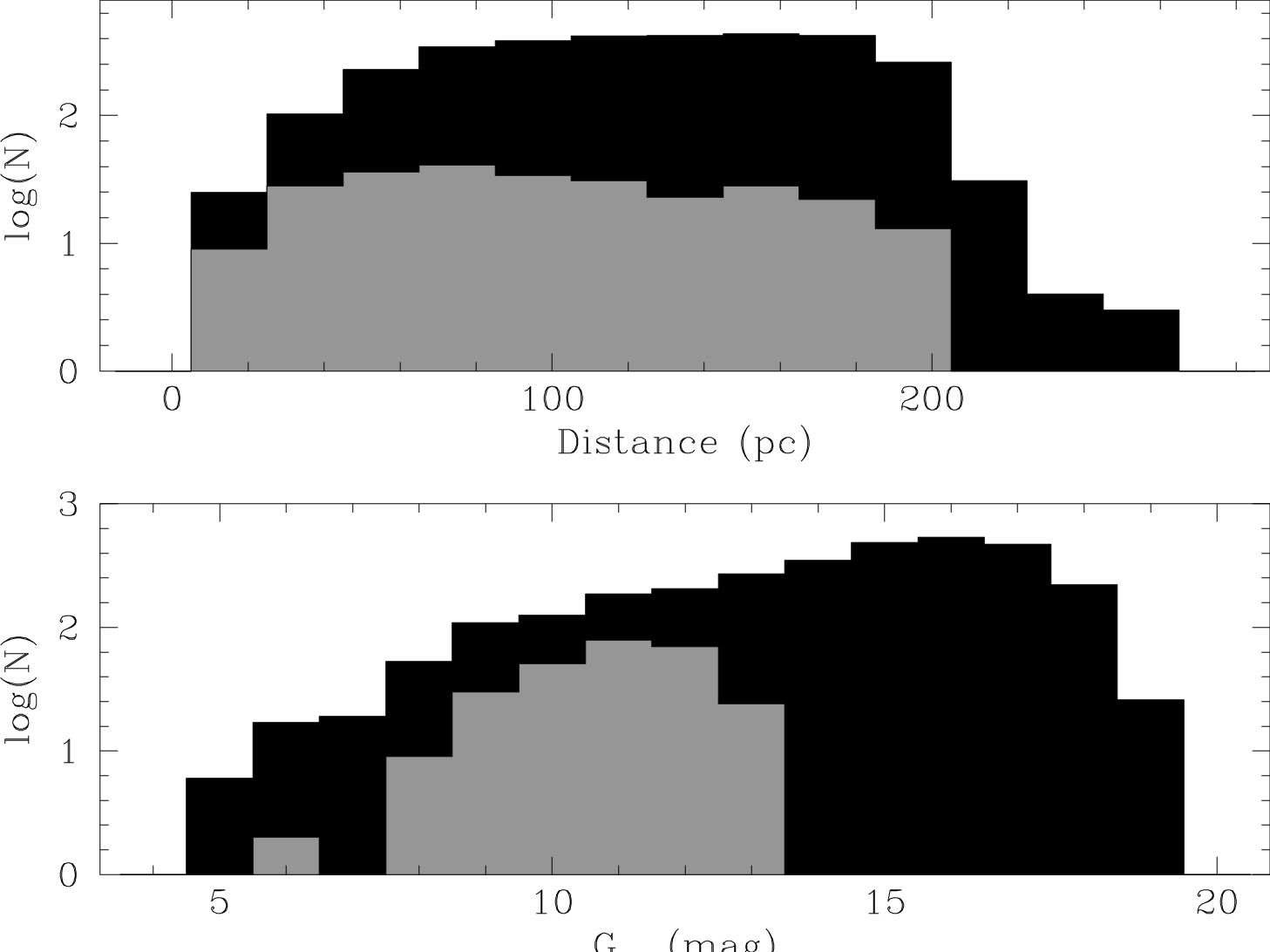}
    \caption{Distance and  main sequence star \emph{Gaia}  G magnitude
      for      the      entire      sample     of      CPMPs      from
      \citet{rebassa-mansergasetal21} (black) and  for those that were
      followed-up for high-resolution observations (gray).}
    \label{fig:magdist}
\end{figure}

\section{The WDMS binary sample}

In  \citet{rebassa-mansergasetal21} we  provided a  catalogue of  4415
common proper motion pairs (CPMPs) formed  by a white dwarf and a main
sequence  star,   identified  within   the  second  data   release  of
\emph{Gaia}  \citep{Gaia2018}.   We   also  performed  high-resolution
spectroscopic observations\footnote{We  used the  Mercator, Telescopio
  Nazionale    Galileo   and    Xinglong    2.16m   telescopes,    see
  \citet{rebassa-mansergasetal21} for details  on the observations and
  their data  reduction.} of  235 main  sequence companions  to derive
their [Fe/H]  abundances, which together  with the total  ages derived
from  the   white  dwarf  primaries   allowed  us  to   constrain  the
age-metallicity relation  in the  solar neighbourhood. Given  that the
primary aim of  that work was to measure the  [Fe/H] abundances of the
companions, our  follow-up observations focused  on F, G, K  and early
type M dwarfs, thus excluding M dwarfs of effective temperatures under
$\simeq$3000\,K for  which it becomes extremely  challenging to derive
metallicites.  Thus,  the present  sample of observed  targets clearly
lacks  objects  containing  fully convective  (intrinsically  fainter)
companions and comprises  objects that are brighter and  closer to the
Sun  as  compared   to  the  entire  catalogue  of   4415  CPMPs  (see
Figure\,\ref{fig:magdist}).  To this  sample of 235 WDMS  CPMPs we add
in  this  work 427  pairs  that  were  either  part of  our  follow-up
high-resolution   observations   but    have   no   [Fe/H]   abundance
determinations  due  to the  lack  of  convergence  in the  fits  (115
objects) or  had available  low/medium resolution spectra  from LAMOST
(Large      Area       Multi-Object      Spectroscopic      Telescope;
\citealt{Cui2012RAA}) data  release 8  (312 objects). Thus,  our total
initial sample of  study in this work consisted of  662 WDMS CPMPs for
which  the main  sequence companions  have available  spectra obtained
from our  high-resolution spectroscopic follow-up  observations and/or
from LAMOST DR8.

\begin{figure}
    \centering
    \includegraphics[angle=-90,width=\columnwidth]{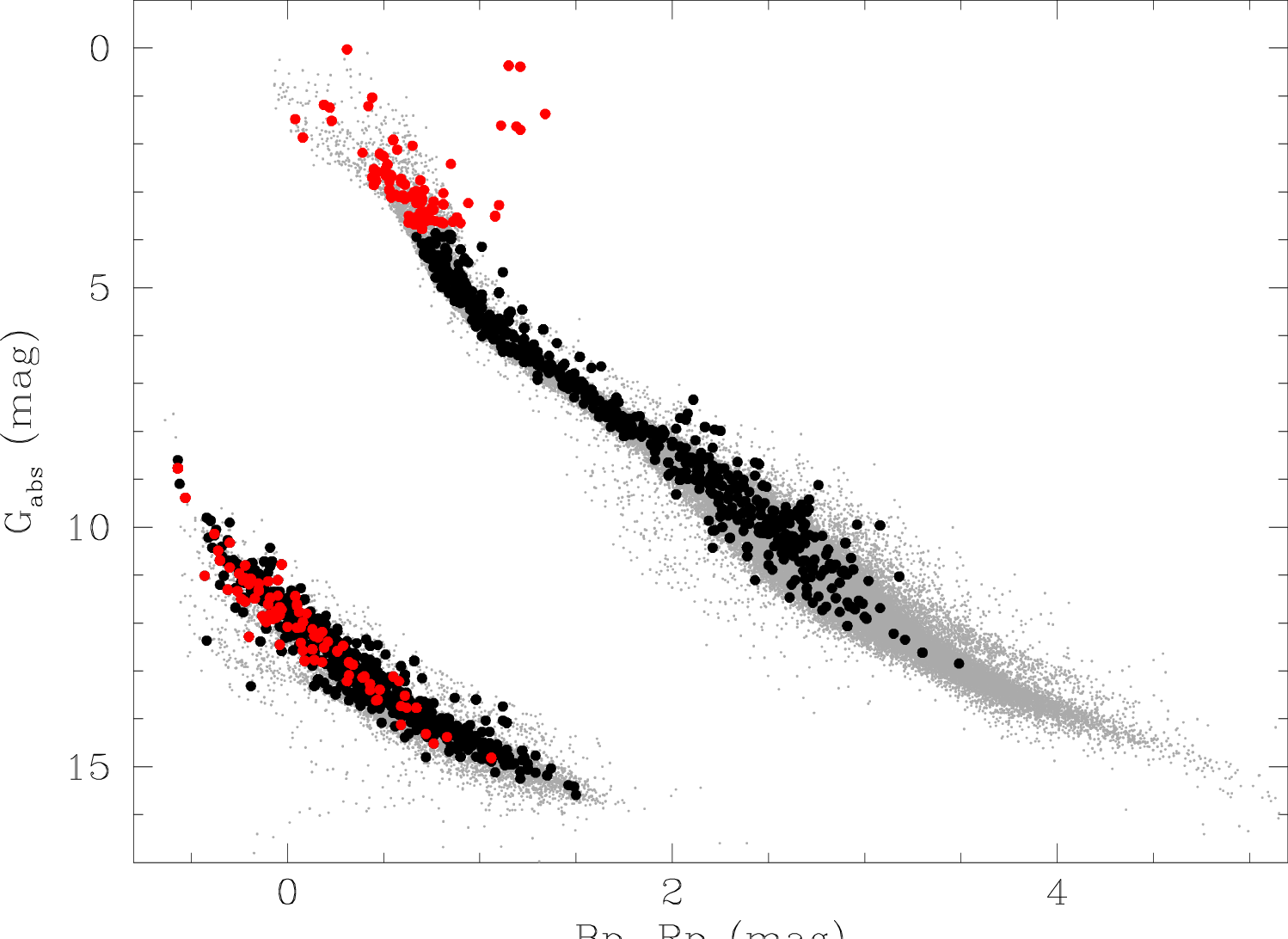}
    \caption{\emph{Gaia}  G$_\mathrm{abs}$  vs.    Bp-Rp  diagram  for
      single white dwarfs and main  sequence stars within 100 pc (gray
      dots), for  our initial WDMS  CPMP sample (red plus  black solid
      dots)  and for  our final  sample to  be analysed  in this  work
      (black  solid dots)  after  excluding 85  giant  stars and  main
      sequence stars above the Kraft break (red solid dots).}
    \label{fig:HR}
\end{figure}

\begin{figure*}
    \centering
    \includegraphics[width=0.7\textwidth]{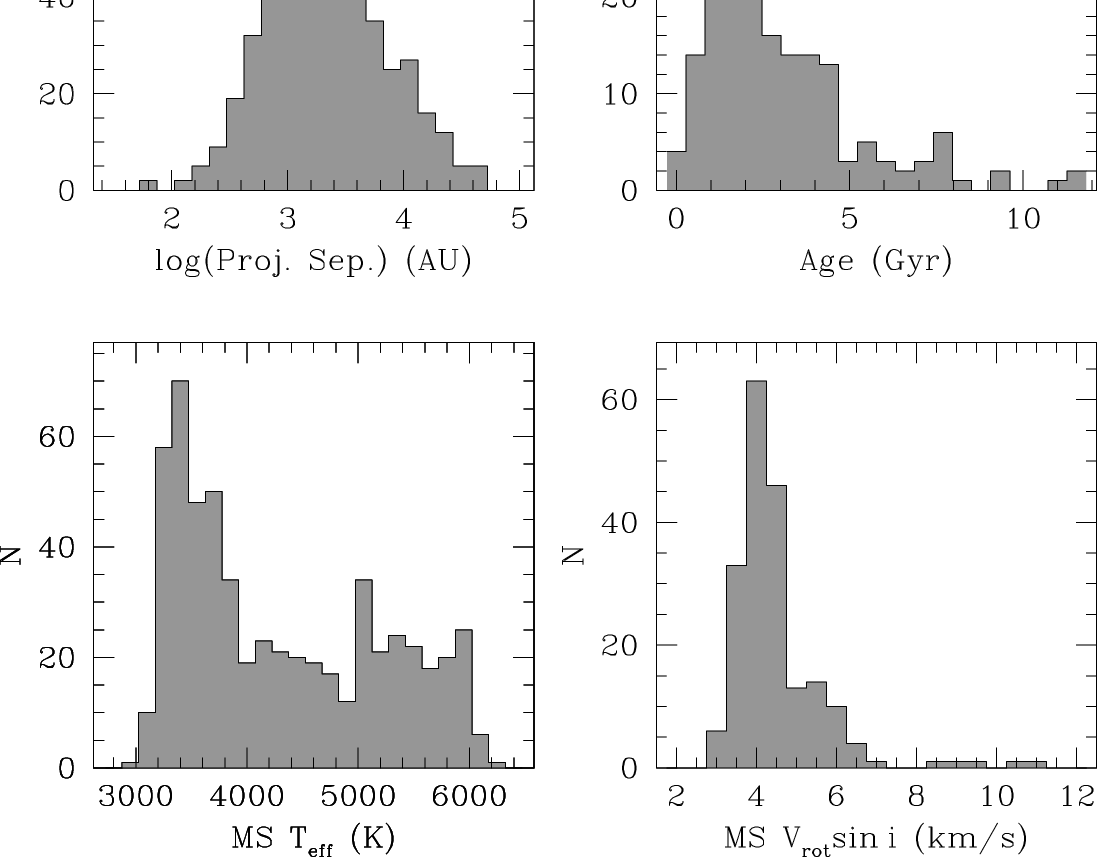}
    \caption{Distribution of projected orbital separations (top-left),
      ages (top-right;  displaying only those values  with errors less
      than 1\,Gyr), main sequence effective temperatures (bottom-left)
      and rotational velocities (bottom-right; when available).}
    \label{fig:distr}
\end{figure*}

The \textit{Gaia} M$_\mathrm{G}$ vs.   Bp-Rp diagram for these targets
is  illustrated in  Figure\,\ref{fig:HR}.   Inspection  of the  figure
reveals that  a few giant stars  are included in our  sample, and also
the presence of stars above the  Kraft break (i.e.  the limit of outer
convective  envelope  and   occurrence  of  significant  chromospheric
activity;  \citealt{Kraft1967}),  which  corresponds to  an  effective
temperature  of  $\simeq$6200\,K,   or  M$_\mathrm{G}  \simeq$3.8  mag
\citep{Pecaut2013}.   Therefore, we  excluded these  85 giant  and hot
main   sequence   stars   by  simply   considering   companions   with
M$_\mathrm{G}<3.8$ mag,  thus reducing our  total sample to  574 CPMPs
(see Table\,\ref{t-summary}).  The coordinates of the 574 objects, the
telescope used  in each  case and  the parameters  obtained throughout
this  work are  provided in  a supplementary  table, available  in the
online version of the paper.

\subsection{Orbital separations}
\label{s-separ}

As  we have  already mentioned  in the  introduction, the  possibility
exists that the  progenitors of the white dwarfs in  our selected WDMS
binaries transferred material to the secondary stars via stellar winds
in the  past, when the  orbital separations were shorter.   This could
have increased the rotation of the current main sequence companions to
the    white     dwarfs,    therefore    affecting     the    expected
age-rotation-activity    relations    \citep{Zurlo2013,    Leiner2018,
  Bowler2021, Gratton2021}.  To explore this possibility, we estimated
the  projected  orbital  separations  of  our  574  CPMPs  from  their
\emph{Gaia}  positions,  which are  shown  in  the top-left  panel  of
Figure\,\ref{fig:distr}.   Assuming  that   the  orbits  adiabatically
expanded   by   a   maximum    factor   of   $\rm   {M_{prog}/M_{WD}}$
\citep{Jeans24} -- where  M$_\mathrm{WD}$ is the white  dwarf mass and
M$_\mathrm{prog}$  its progenitor  mass --  we expect  minimum initial
orbital separations  2-5 times shorter  (depending on the  white dwarf
mass)  than the  current ones,  according to  typical initial-to-final
mass  relations \citep[e.g.][]{Catalan08,  Cummings2018}.  Given  that
there is  a non-negligible  fraction of  systems with  current orbital
separations up  to $\simeq$1000 AU,  we conclude  it is likely  that a
fraction ($\simeq$2--8  per cent) of  WDMS binaries in our  sample may
have suffered from wind accretion episodes in the past.

\subsection{Total ages}
\label{s-ages}

Given  that not  all the  white  dwarfs in  our sample  of study  have
available spectra,  we derived  the total ages  in an  homogeneous way
from their \emph{Gaia} photometry.  The ages of the 235 DR2 WDMS CPMPs
with  available [Fe/H]  abundances  from the  companions were  already
derived by \citet{rebassa-mansergasetal21}.  The method used basically
interpolates the  white dwarf \emph{Gaia} $G$  absolute magnitudes and
Bp-Rp  colours in  the evolutionary  sequences of  the La  Plata group
\citep{Althaus2015, Renedo2010,  Camisassa2016, Camisassa2019}  at the
corresponding  metallicity (or  [Fe/H] abundance)  to derive  both the
white dwarf cooling ages and their main sequence progenitor lifetimes.
This approach of  deriving total ages as the white  dwarf cooling plus
main    sequence   progenitor    lifetime   was    tested   by    e.g.
\citet{Garcia-Berro2013, Torres2015,  Qiu2021}, who derived  the total
ages of white  dwarfs in open and globular clusters  and found them to
be consistent to those obtained from the main sequence turn-off of the
clusters.   Moreover,  using double  white  dwarfs  or triple  systems
containing at  least two white dwarfs,  \citet{Heintz2022} report that
total white  dwarf ages are  generally reliable, especially  for those
with masses above 0.63\,M$_{\odot}$.  This is in line with the results
of  \citet{Fouesneau2019} and  \citet{Moss2022}, who  calculated white
dwarf total ages in the same way as  in this work and found them to be
$\simeq$5-20 per cent precise.  Unfortunately, this does not take into
account  possible  uncertainties  due  to  the  initial-to-final  mass
relation, which may introduce  differences ranging from $\simeq$0.2 to
$\simeq$2\,Gyr  (see Figure  2 of  \citealt{Rebassa2016b}).  To  avoid
this issue a sample of high-mass  white dwarfs in which the progenitor
lifetime is considerably shorter than the cooling age is required.

In \citet{rebassa-mansergasetal21}  we assumed  that all  white dwarfs
had hydrogen-rich atmospheres (DA white dwarfs) and we took extinction
into account using the 3D \emph{Stilism} maps of \citet{Lallement2014}
and \citet{Capitanio2017}.   Here, we used  the same method  to derive
the ages with  the difference that we searched  for available spectral
types of  the white  dwarfs in  our sample.   The spectral  types were
adopted as hydrogen-rich  if the probability to be a  DA measured from
\citet{Jimenez2023} for  white dwarfs  within 100\,pc was  larger than
0.5.   For  those  with  a  probability  under  0.5,  we  adopted  the
classification   based    on   a   random   forest    algorithm   from
\citet{Garcia-Zamora2023},  which  was  developed  to  classify  these
(non-DA)  systems within  100\,pc  into  DBs, DCs,  DQs  and DZs.   We
complemented the  search, especially  for white dwarfs  with distances
above   100\,pc,    using   the   Montreal   White    Dwarf   Database
\citep{Dufour2017}.  Thus,  our sample comprises  220 DAs, 15  DBs, 58
DCs, 9 DQs, 4 DZs and 268 unclassified white dwarfs.  We have employed
the  hydrogen-rich evolutionary  models  of \citet{Camisassa2016}  and
\citet{Camisassa2019} for DA white dwarfs and all DC white dwarfs with
Bp-Rp  colours larger  than 0.8  mag (which  corresponds to  effective
temperatures  under   $\simeq$5000\,K),  the   pure  helium-atmosphere
evolutionary models of \citet{Camisassa2017}  for DB white dwarfs, and
the   white   dwarf   models  with   helium-atmospheres   and   carbon
traces\footnote{Note that  the carbon  traces are  not visible  in the
  optical but  are expected at ultraviolet  wavelengths.}  that follow
the  so-called "carbon  sequence" enrichment  of \citet{Camisassa2023}
for DQ white dwarfs and for  all DC white dwarfs with Bp-Rp$<0.8$ mag.
No ages were calculated for those white dwarfs with DZ spectral types.
White   dwarfs  with   no   classification  were   considered  to   be
hydrogen-rich.  There  is a possibility  that non-DAs exist  among the
268   unclassified  objects.    They   have   Bp-Rp  colours   between
$\simeq-0.3$  and  $\simeq$0.9  mag,  which  correspond  to  effective
temperatures  between $\simeq$20\,000  and $\simeq$5\,000  K for  both
hydrogen-rich  and hydrogen-deficient  white dwarfs  according to  the
cooling  sequences.  The  fraction of  non-DA white  dwarfs goes  from
$\simeq10$  per  cent  at  20\,000\,K to  $\simeq25-35$  per  cent  at
10\,000-5\,000\,K \citep{McCleery2020, Lopez2022, Torres2023}.  Taking
these values into  consideration, we estimate that around  30 per cent
of  the  unclassified  objects   ($\simeq$85)  are  probably  non-DAs.
According  to \citet{Camisassa2017},  cooling age  differences between
hydrogen-rich and  helium-rich white  dwarfs are lower  than 0.2\,Gyr,
with hydrogen-rich  white dwarfs being  younger for a  given effective
temperature.  The age discrepancy may be further accentuated since the
derived white dwarf masses slightly  vary when using hydrogen-rich and
helium-rich models, which implies the progenitor lifetimes change too.
If the white  dwarfs are of low mass, the  total age discrepancies can
be as high as 5--6 Gyr. This is the case when using DQ models too.

It has to  be emphasised that the evolutionary sequences  we have used
do not provide total ages for  white dwarfs that are less massive than
$\simeq0.5$\,M$_{\sun}$,  since  the   progenitors  of  these  objects
require ages  longer than the  Hubble time to  evolve out of  the main
sequence and in this case the use of an initial-to-final mass relation
is not  appropriate.  Thus, for 432  of the 574 objects  we managed to
derive an  age and, among  these, 201  have age uncertainties  of less
than 1\,Gyr.  Since  the main motivation of this work  is to constrain
the  age-rotation-activity relation,  we only  considered the  ages of
these 201 objects in our analysis.  Note that of these 201, 85 have no
white  dwarf type  classification,  therefore we  expect  the ages  of
$\simeq$30 per cent of those  ($\simeq$25 objects; i.e. $\simeq$12 per
cent of the full sample with ages) not to be fully reliable, as stated
above.  The resulting age distribution is illustrated in the top-right
panel of  Figure\,\ref{fig:distr}, where it  can be seen that  most of
the systems have ages concentrated in the 1-4\,Gyr bins.

Finally, it is  also important to note that for  the objects that were
not  studied in  \citet{rebassa-mansergasetal21}  we  adopted a  solar
abundance  for the  age calculation  since the  metallicites in  these
cases  were not  known.   According to  the  white dwarf  evolutionary
models used  in this  work, which employ  the progenitor  lifetimes of
\citet{Miller2016}\footnote{These   main    sequence   lifetimes   are
  consistent to those of  \citet{Pietrinferni2004} and the theoretical
  initial-to-final massa  relation employed  is remarkably  similar to
  the one  by \citet{Catalan08} for Z=0.02  and similar to the  one by
  \citet{Cummings2018}  for  Z=0.001.},   this  assumption  may  imply
progenitor  lifetime (hence  total age)  differences respect  to those
obtained   assuming   the   solar   metallicity   value   of   up   to
$\simeq$1.5\,Gyr        for        low-mass        white        dwarfs
($\simeq$0.55\,M$_{\odot}$),   with  the   average  difference   being
0.15$\pm$0.4\,Gyr over  the entire  white dwarf mass  range.  However,
note that  only 14  of the  201 white  dwarfs with  age determinations
considered in this work have  masses below 0.57\,M$_{\odot}$ and, as a
consequence, we  do not expect  this issue to considerably  affect our
results.  Regarding the total age errors, these were obtained directly
from  considering the  parallax (hence  absolute magnitude)  and Bp-Rp
uncertainties.

\subsection{Main sequence star effective temperatures}
\label{s-teffs}

The main sequence stars in our CPMPs were observed by a wide number of
different telescopes  and spectroscopic  resolutions, and  the spectra
were subject to different signal-to-noise ratios. As a consequence, in
order to  acquire an as  homogeneous as possible compilation,  we used
the $Gaia$  photometry of  each star  rather than  their corresponding
spectra  to derive  their  effective  temperatures.  We  independently
interpolated the extinction corrected  $G$ absolute magnitudes and the
Bp-Rp      colours       in      the      updated       tables      of
\citet{Pecaut2013}{\footnote{\url{https://www.pas.rochester.edu/~emamajek/EEM_dwarf\_UBVIJHK\_colors\_Teff.txt}}
  to derive two effective temperature values for each star, which were
  then averaged to get a  single measurement (the error was considered
  as  the  standard  deviation).  The  corresponding  distribution  is
  provided in  the bottom-left  panel of  Figure\,\ref{fig:distr}.  As
  expected,  no  main  sequence  stars can  be  found  with  effective
  temperatures higher  than $\simeq$6200\,K, which corresponds  to the
  Kraft  break, since  these objects  were excluded  from our  sample.
  Moreover, the  majority of  stars have effective  temperatures above
  3000\,K,  as   a  consequence   of  our   high-resolution  follow-up
  observations prioritizing F, G, K and early M dwarf companions.

\subsection{Main sequence star rotational velocities}
\label{s-rot}

The    Mercator,    TNG    and   Xinglong    spectra    we    obtained
\citep{rebassa-mansergasetal21} are of sufficient resolution to derive
projected  rotational velocities.  To  that end  we  used the  Fourier
transform                        (FT)                        technique
\citep[e.g.][]{2008oasp.book.....G}. Following  \cite{Maldonado22}, we
used  three   spectral  lines   at  6335.33\,\AA,   6380.75\,\AA,  and
6393.61\,\AA\  for the  computations.   The values  obtained from  the
different lines  were averaged  into a  single measurement,  while the
standard deviation  was considered as the  associated uncertainty. The
dominant term  in the FT  of the  rotational profile is  a first-order
Bessel function that produces a series of relative minima at regularly
spaced  frequencies.  The  first  zero  of the  FT  and the  projected
rotational velocity are related by

 \begin{equation}
  v_\mathrm{rot}\sin i=\frac{c}{\lambda}\times\frac{k_{1}}{\sigma_{1}}
 \end{equation}

\noindent where $\lambda$ is the  central wavelength of the considered
line, $c$ is the speed of light, ${\sigma_{\rm 1}}$ is the position of
the first zero of the FT, and  ${k_{\rm 1}}$ is a function of the limb
darkening  coefficient ($\epsilon$).  The  term ${k_{\rm  1}}$ can  be
approximated     by      a     fourth-order      polynomial     degree
\citep{1990A&A...237..137D}

 \begin{equation}
 k_{1} = 0.610 + 0.062\epsilon + 0.027\epsilon^{2} + 0.012\epsilon^{3}
 + 0.004\epsilon^{4}
 \end{equation}

\noindent   where    we   assume    $\epsilon$   =    0.6   \citep[see
  e.g.][]{2008oasp.book.....G}.

We managed to derive projected rotational velocities for 195 stars and
the corresponding  distribution is provided in  the bottom-right panel
of Figure\,\ref{fig:distr}. It  reveals that the vast  majority of the
main  sequence  stars  in  our sample  are  apparently  slow  rotators
(V$_\mathrm{rot}$sin\,i$<$6\,km/s)  and  that the  minimum  detectable
value is  $\simeq$3\,km/s.  The typical  errors are between 0.1  and 2
km/s with an average  value of $\simeq1.2\pm1$\,km/s, independently of
the spectrograph used.

\begin{figure}
    \centering
    \includegraphics[angle=-90,width=\columnwidth]{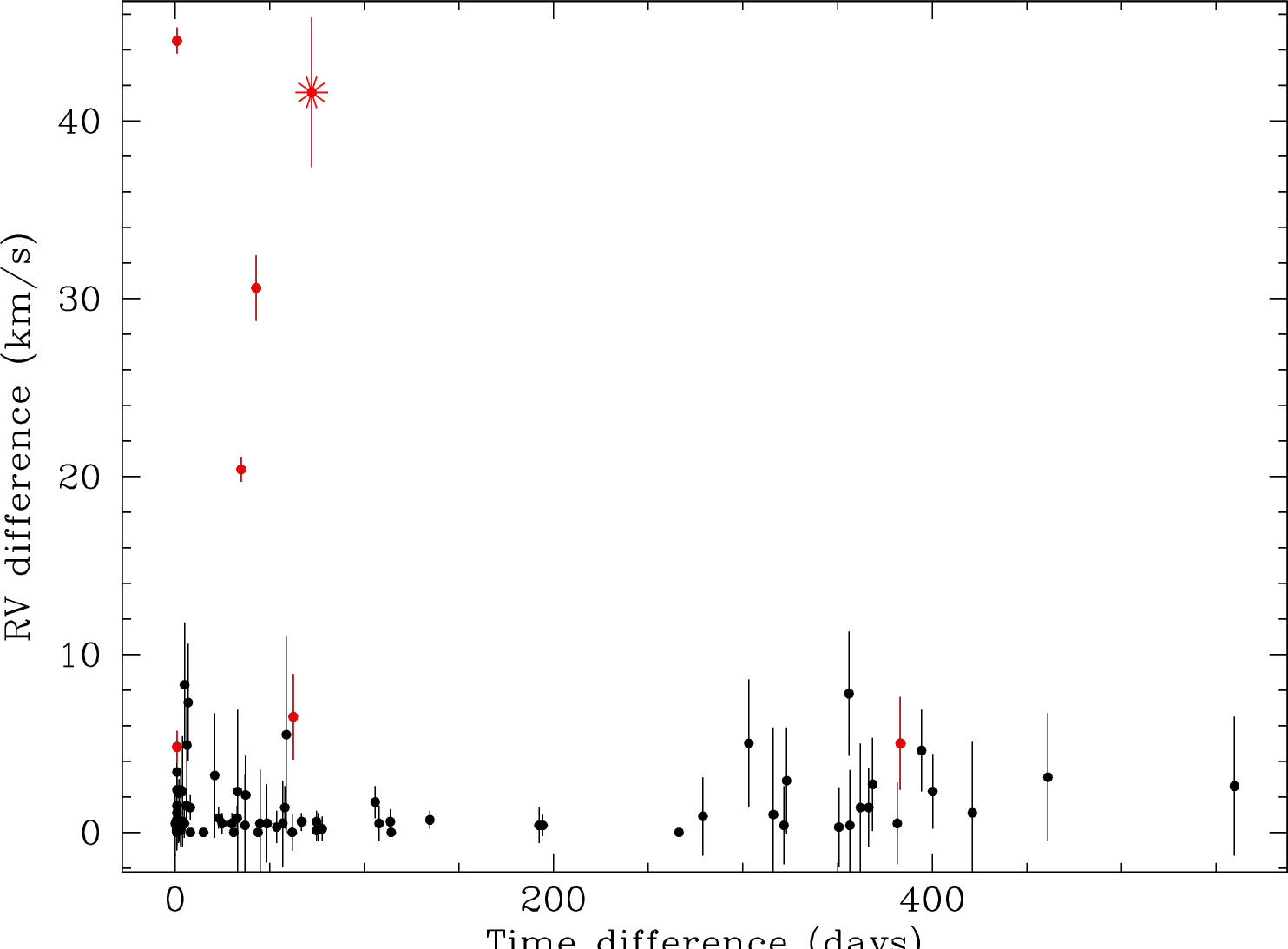}
    \caption{The absolute value of the radial velocity difference as a
      function of  the absolute  value of  the difference  between the
      heliocentric  Julian  dates  corresponding to  the  maximum  and
      minimum radial velocities.  Red solid  dots indicate the 7 stars
      displaying more  then 3$\sigma$  radial velocity  variation. The
      red star indicates this object displays H$\alpha$ in emission.}
    \label{fig:rvdif}
\end{figure}

\begin{figure*}
    \centering
    \includegraphics[angle=-90,width=0.7\textwidth]{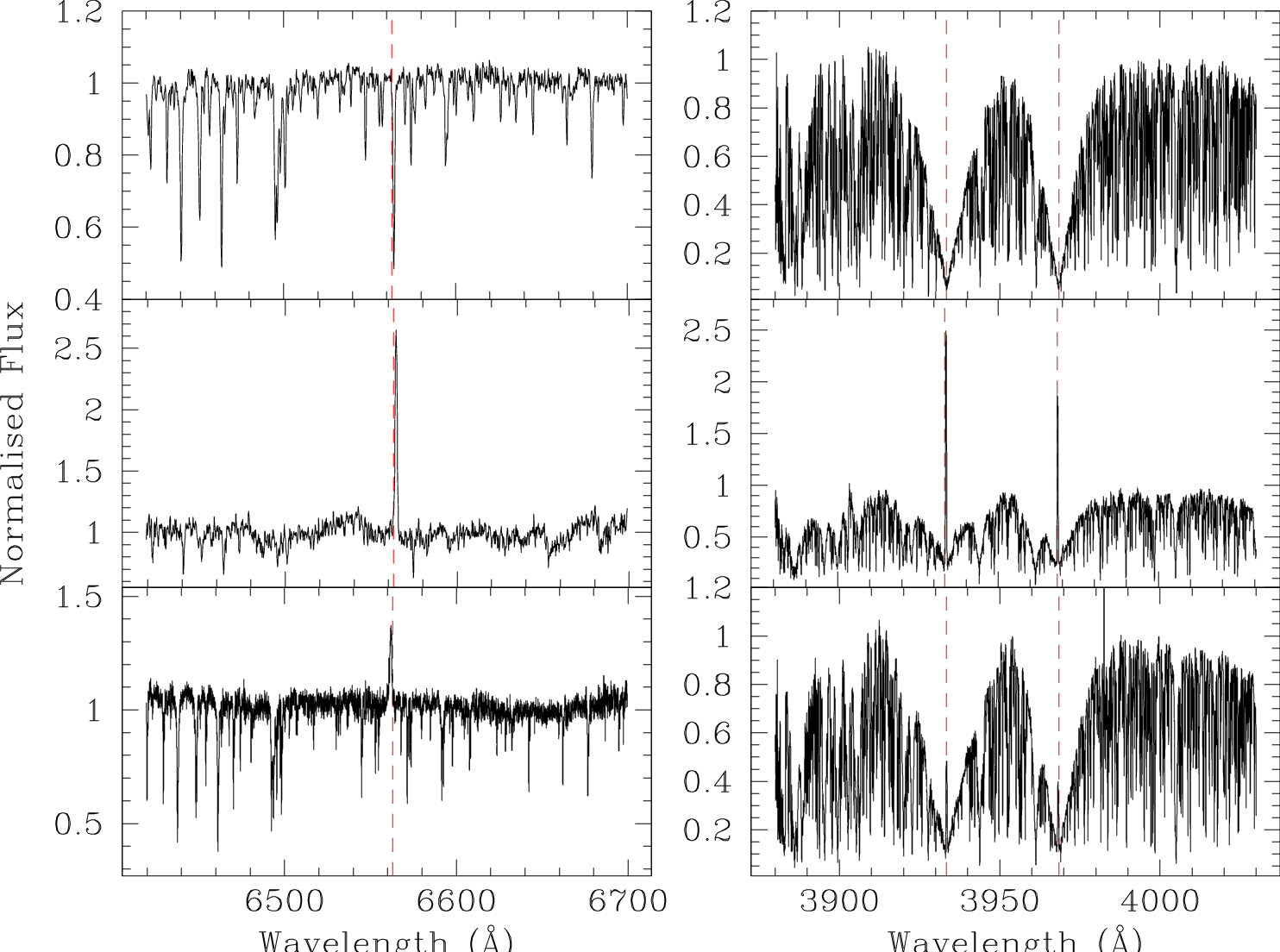}
    \caption{Left panels:  example spectra of inactive  (top; spectrum
      of  J1217+3423 obtained  by  LAMOST with  the medium  resolution
      instrument), active (middle; spectrum  of J0001+3559 obtained by
      LAMOST with the medium  resolution instrument) and weakly active
      (bottom;   J1014+6155,  considered   as   active  after   visual
      inspection, obtained  by the TNG, Telescopio  Nazionale Galileo,
      with the HARPS-N spectrograph) objects  based on the presence or
      lack of  H$\alpha$ emission. Right panels:  example spectra (all
      obtained with  the Mercator  telescope equipped with  the HERMES
      spectrograph)  of inactive  (top;  J2328+1045), active  (middle;
      J0221+5333) and weakly active (bottom; J0209-0140, considered as
      active after visual inspection) objects based on the presence or
      lack  of \Ion{Ca}{ii}~H\&K  emission.  The  dashed red  vertical
      lines   indicate    the   position   of   the    H$\alpha$   and
      \Ion{Ca}{ii}~H\&K lines.}
    \label{fig:spec}
\end{figure*}

\subsection{Main sequence star radial velocities}

Emission lines (e.g. H$\alpha$) due to magnetic activity may arise not
only for main sequence stars that are young in our sample, but also if
they are in  close orbits with unseen companions  and therefore forced
to rotate fast  due to tidal locking.  It that was the  case, we would
detect radial velocity variations.

We derived the radial velocities  from the TNG, Mercator, Xinglong and
LAMOST  medium resolution  spectra of  the main  sequence stars  in an
homogeneous way following \citet{Rebassa2007}: we fitted the H$\alpha$
emission/absorption with a single Gaussian plus a parabola.  H$\alpha$
is the  only suitable line  sampled by  the spectra obtained  from all
telescopes. For  287 of the 574  systems in our sample  we measured at
least  one  radial velocity;  for  107  two  or  more.  For  105,  the
available  velocities come  from  spectra separated  by  at least  one
night.  We detected more than 3$\sigma$ radial velocity variation in 7
systems  ($\simeq$6  per  cent  of  the  sample  with  more  than  one
velocity). For the 107 objects with more than one measurement, we show
in Figure\,\ref{fig:rvdif}  the absolute value of  the radial velocity
difference as a function of the  absolute value of the time difference
between  the  heliocentric Julian  date  of  the minimum  and  maximum
velocities.

The 7 stars displaying more than  3$\sigma$ variation are shown in red
in   Figure\,\ref{fig:rvdif}.   These   are  J0227+6355,   J0309+3340,
J0314+3248,  J0708+7831, J0804+1207,  J1514+0157 and  J1559+2528. Only
J0708+7831,  associated to  a  time difference  of  72 days,  displays
H$\alpha$ in  emission (none display \Ion{Ca}{ii}~H\&K  emission).  We
do not derive a reliable  age for this object (Section\,\ref{s-ages}),
therefore the emission might be a simple consequence of the star being
young.  The current projected orbital  separation is estimated as 1991
AU, indicating that  it is unlikely that wind accretion  took place in
the past.  The remaining 6 objects display H$\alpha$ in absorption and
are all G  and K dwarfs with projected orbital  separations above 1600
AU, with the exception of J1514+0157 which has a separation of 680 AU.
Except perhaps  for J1514+0157,  the radial velocity  variations might
arise due to the presence of  hidden, nearby and less luminous M dwarf
companions.   That  is, they  are  candidates  for being  hierarchical
triple systems  composed of an  inner unresolved main  sequence binary
and an  outer white dwarf.  Note  that this is also  a possibility for
J0708+7831, which displays H$\alpha$ in emission.

\subsection{Activity indicators}
\label{sec:indicators}

We used two activity indicators  to evaluate whether the main sequence
stars in our WDMS CPMPs are  magnetically active or not. The first was
the presence  of H$\alpha$  emission, through  the measurement  of the
equivalent  width  of  the  line.   The second  was  the  presence  of
\Ion{Ca}{ii}~H\&K emission,  via the  analysis of  the $S_\mathrm{HK}$
index.

We  measured the  H$\alpha$  equivalent width  (EW$_\mathrm{H\alpha}$)
following the  method described in Paper  I. That is, we  first fitted
the flux in the 6420--6700\,\AA\ range with a parabola. The fit, which
excluded  the H$\alpha-7  < \lambda  <$ H$\alpha+7$\,\AA\  region, was
then  used to  normalise  the spectra.  The EW$_\mathrm{H\alpha}$  was
measured from the  normalised spectra within the  H$\alpha-7 < \lambda
<$ H$\alpha+7$\,\AA\ region and active  stars were identified from the
following criteria:

\begin{eqnarray}
\mathrm{SNR}\geq10\\
\mathrm{EW}_{\mathrm{H}\alpha} \leq -0.75\,\mbox{\AA} \\
|\mathrm{EW}_\mathrm{H\alpha}| > 3 \times |\mathrm{eEW}_\mathrm{H\alpha}|\\
\mathrm{h} > 3 \times N_\mathrm{cont}
\end{eqnarray}

\noindent
where  SNR   is  the  signal-to-noise   ratio  of  the   spectra,  the
EW$_\mathrm{H\alpha}$ error is eEW$_\mathrm{H\alpha}$, $h$ is the peak
of the flux at H$\alpha$ above the continuum, and $N_\mathrm{cont}$ is
the    noise    at    continuum    level   (see    an    example    in
Figure\,\ref{fig:spec},  middle left  panel). We  flagged as  inactive
stars those fulfilling equation 1  and the inverted forms of equations
2--4 (see an example in Figure\,\ref{fig:spec}, top left panel). Those
stars that did not pass the active or inactive criteria (mainly due to
the low SNR of their spectra) were not given an activity index.

Main  sequence  stars  with  more than  one  available  spectrum  were
considered  inactive when  no  emission  was detected  in  any of  the
spectra.   Thus, the  above exercise  resulted  in 33  active and  525
inactive stars according  to the detection of  H$\alpha$ emission. The
remaining 16  objects had  all their  spectra of SNR  under 10  and we
hence  decided  not  provide   any  activity  classification.   Visual
inspection of  the spectra revealed  that only $\simeq$3  per\,cent of
the sample was misclassified.  In  particular, 17 spectra were flagged
as inactive whilst visual  inspection revealed mild H$\alpha$ emission
(see an example in Figure\,\ref{fig:spec}, bottom left panel).

The $S_\mathrm{HK}$ index is our second activity indicator, especially
useful for F, G  and K (hereafter FGK) stars since it  is based on the
\Ion{Ca}{ii}~H\&K doublet. It is defined as follows \citep{Wilson1968,
  Wilson1978, Vaughan1978, Duncan1991, Baliunas1995}:
\begin{equation}
S_\mathrm{HK} = \beta \frac{\mathrm{H+K}}{\mathrm{R+V}}
\end{equation}
\noindent
where H and  K are the integrated  fluxes (or number of  counts if the
spectra are  not flux-calibrated)  within 2\,\AA\  rectangular windows
around  the  \Ion{Ca}{ii}~H  ($3968.47\pm1$\,\AA)  and  \Ion{Ca}{ii}~K
($3933.66\pm1$\,\AA) lines,  respectively; R and V  are the integrated
fluxes  (or  number of  counts)  within  20\,\AA\ rectangular  windows
(i.e.  continuum  windows) to  the  right  (3991--4011\,\AA) and  left
(3891--3911\,\AA) sides  of the H\&K lines,  respectively, and $\beta$
is calibration constant.  The choice of 2\,\AA\  rectangular boxes for
obtaining H  and K follows \citet{Zhao2011},  which implies $\beta=10$
since  the  continuum  windows  are  10  times  wider  than  the  H\&K
windows.  To measure  H,  K, R,  and V,  the  spectra were  previously
normalised. We considered first the sum of the fluxes at ten continuum
ranges (defined as  3881$\pm$1\,\AA, 3891$\pm$1\,\AA, 3901$\pm$1\,\AA,
3911$\pm$1\,\AA,  3956$\pm$1\,\AA,  3981$\pm$1\,\AA,  3991$\pm$1\,\AA,
4001$\pm$1\,\AA, 4011$\pm$1\,\AA and  4021$\pm$1\,\AA) and fitted then
these  fluxes  with a  second  order  polynomial,  which was  used  to
normalise the spectra.

It has to  be emphasised that the $S_\mathrm{HK}$  index obtained here
should  not be  directly compared  to  that from  the original  survey
definition  \citep{Wilson1968} since  they employed  triangular rather
than rectangular H\&K windows (although  this effect is expected to be
minor, see  \citealt{Hall2007}) and also because  the $\beta$ constant
is different. However, it perfectly  serves our purpose as an activity
indicator.    In  this   sense,  we   selected  all   stars  with   an
$S_\mathrm{HK}$ value greater (smaller) than 0.75 as active (inactive)
stars  (see   examples  in  the   middle  and  top  right   panels  of
Figure\,\ref{fig:spec}, respectively). This  threshold was obtained by
analysing the $S_\mathrm{HK}$  indexes of those stars  that are active
or  inactive based  on the  H$\alpha$  emission of  their spectra  and
resulted in 29 active and 157  inactive stars. Note that the number of
stars with  activity indicators provided by  the $S_\mathrm{HK}$ index
is smaller  than that with  H$\alpha$ emission indicators  because (1)
the   LAMOST   medium   resolution    spectra   do   not   cover   the
\Ion{Ca}{ii}~H\&K  lines  and  (2)  for several  stars,  especially  M
dwarfs, the \Ion{Ca}{ii}~H\&K lines were too noisy.  Visual inspection
of the spectra  revealed that in $\simeq$6 per\,cent of  the cases the
above exercise  resulted in wrong classifications.   In particular, 17
spectra with $S_\mathrm{HK}$ index below 0.75 turned out to be active,
12  from observed  by the  TNG/HARPS-N  and 5  by the  Mercator/HERMES
telescope/spectrograph.   These  two  configurations yield  very  high
resolution spectra (R=115\,000 and 85\,000 respectively), which allows
visually detecting weak chromospheric H\&K emission (see an example in
the bottom right panel of Figure\,\ref{fig:spec}).

After   the   above  exercises,   we   combined   the  H$\alpha$   and
$S_\mathrm{HK}$  indicators  to give  a  final  activity flag  to  our
stars. That  is, if the two  indicators are available we  considered a
star  as inactive  if both  classifications are  inactive. A  star was
considered as active if at least  one of the indicators classified the
star as active. Thus, the final number of active/inactive stars in our
sample  is  49/509, i.e.   an  overall  fraction  of active  stars  of
$\simeq$10  per\,cent  (see   Section\,\ref{s-fractions}  for  further
discussions).

\begin{figure*}
    \centering
    \includegraphics[width=0.49\textwidth]{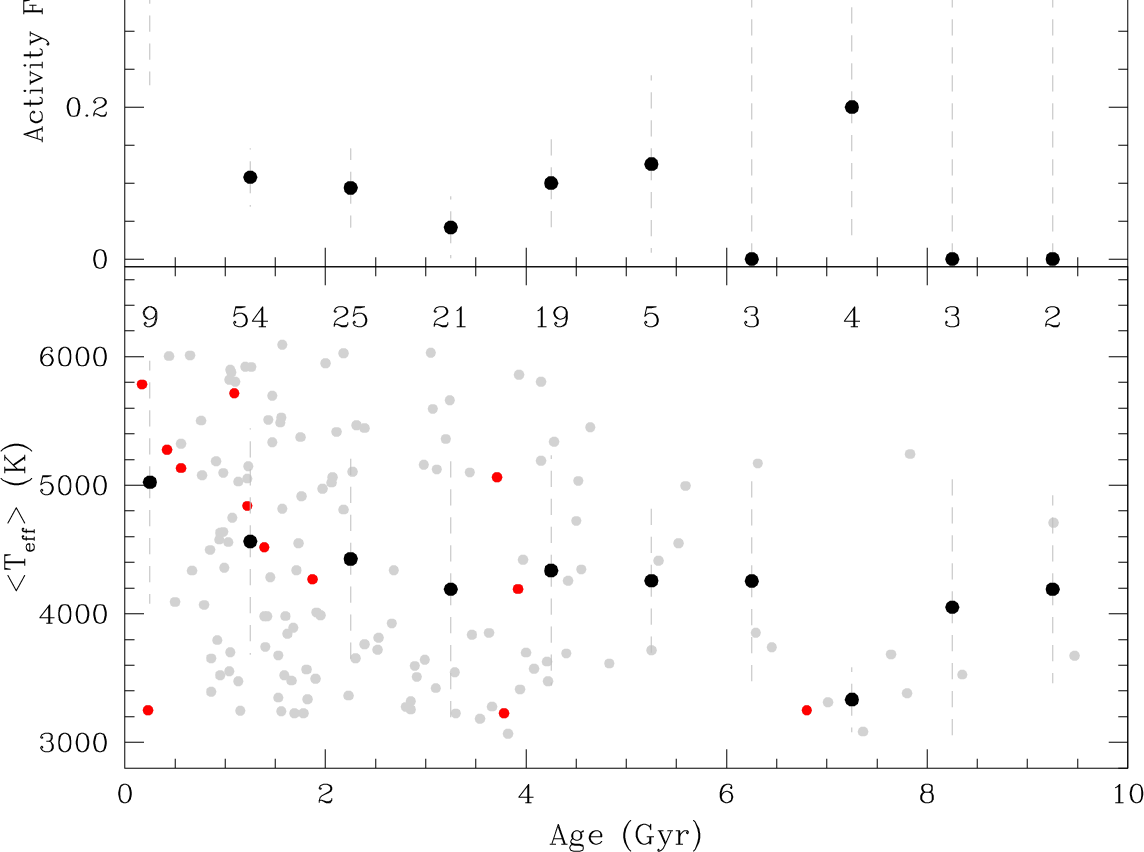}
    \includegraphics[width=0.49\textwidth]{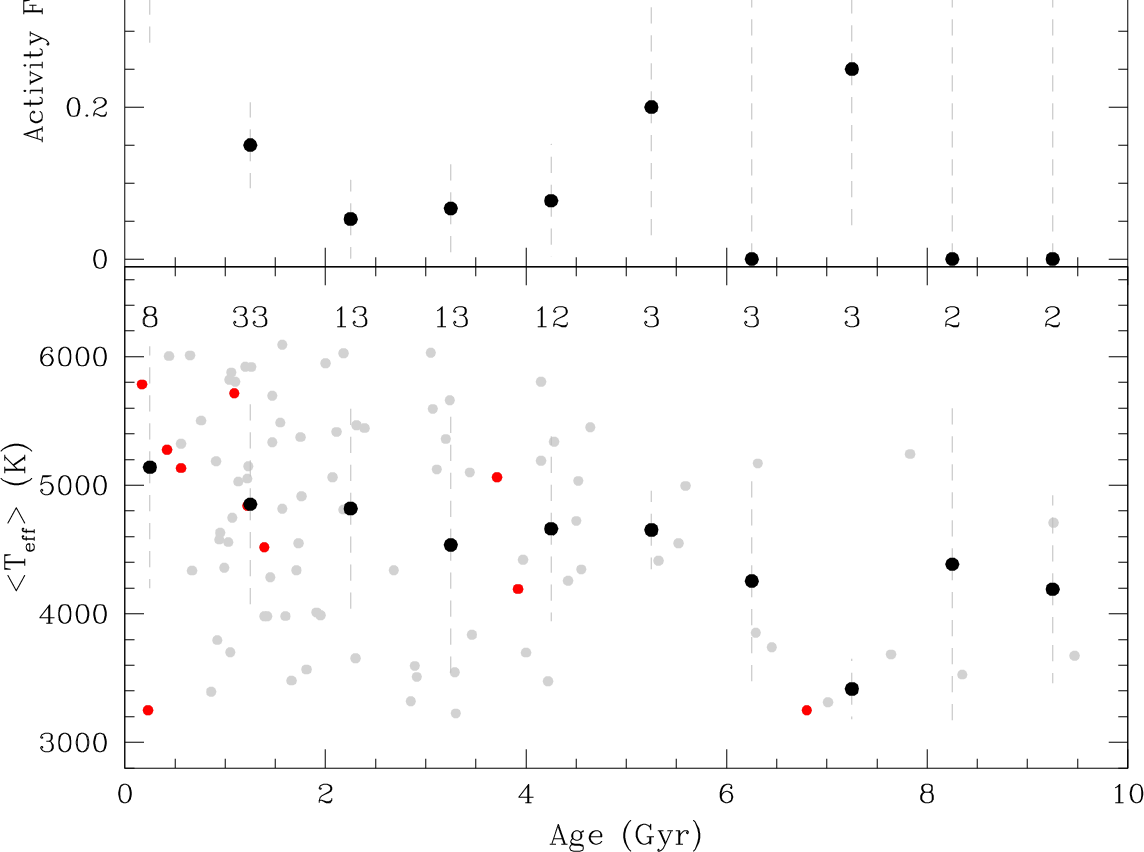}
    \caption{Top panels: activity  fractions as a function  of age for
      the entire  sample of main  sequence companions to  white dwarfs
      studied in  this work (left panel)  and for those stars  that we
      observed with high-resolution  spectrographs (right). The number
      of  stars  in  each  bin  are   indicated  on  the  top  of  the
      panels.   Bottom   panels:   the   distribution   of   effective
      temperatures per  age bin (grey  dots; red dots  indicate active
      stars) together with the associated mean and standard deviations
      (black solid dots and lines,  respectively). The number of stars
      per age  bin are also  indicated in the  top. Note that  not all
      stars  with derived  ages have  measured effective  temperatures
      with uncertainties under 120\,K.}
    \label{fig:age-act1}
\end{figure*}

\begin{figure*}
    \centering
    \includegraphics[width=0.49\textwidth]{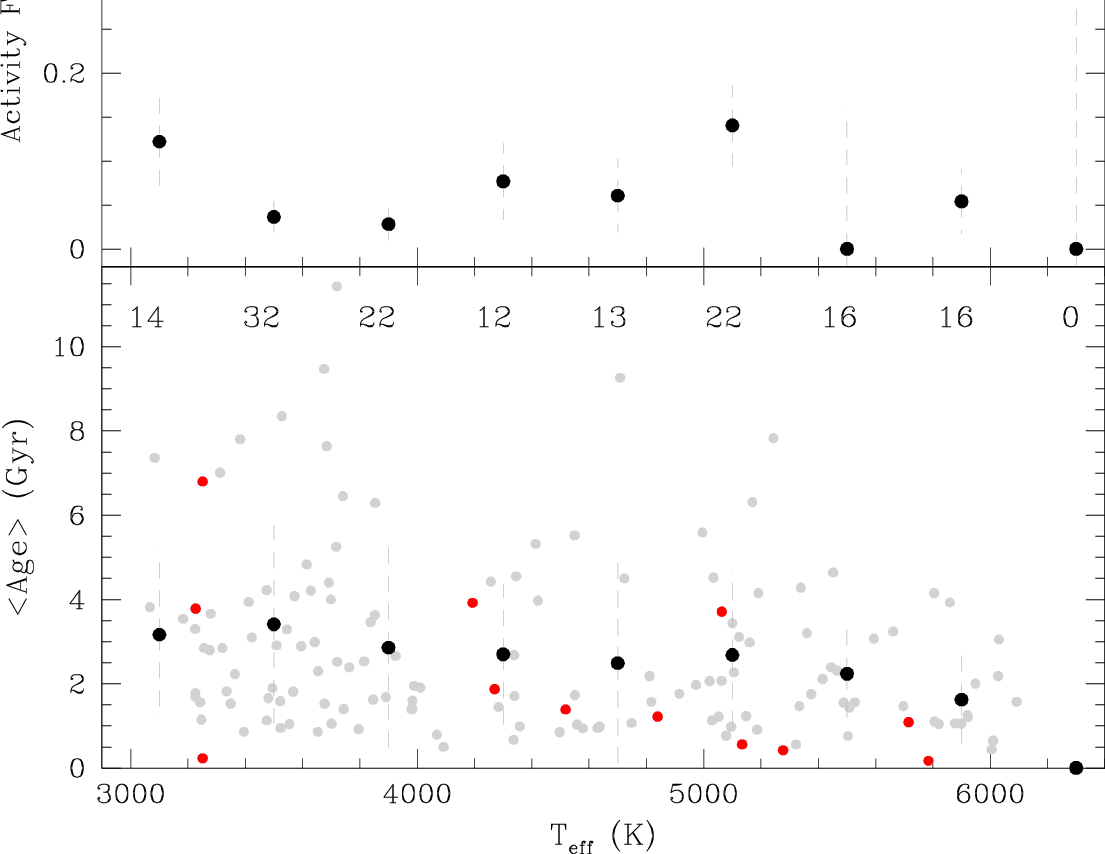}
    \includegraphics[width=0.49\textwidth]{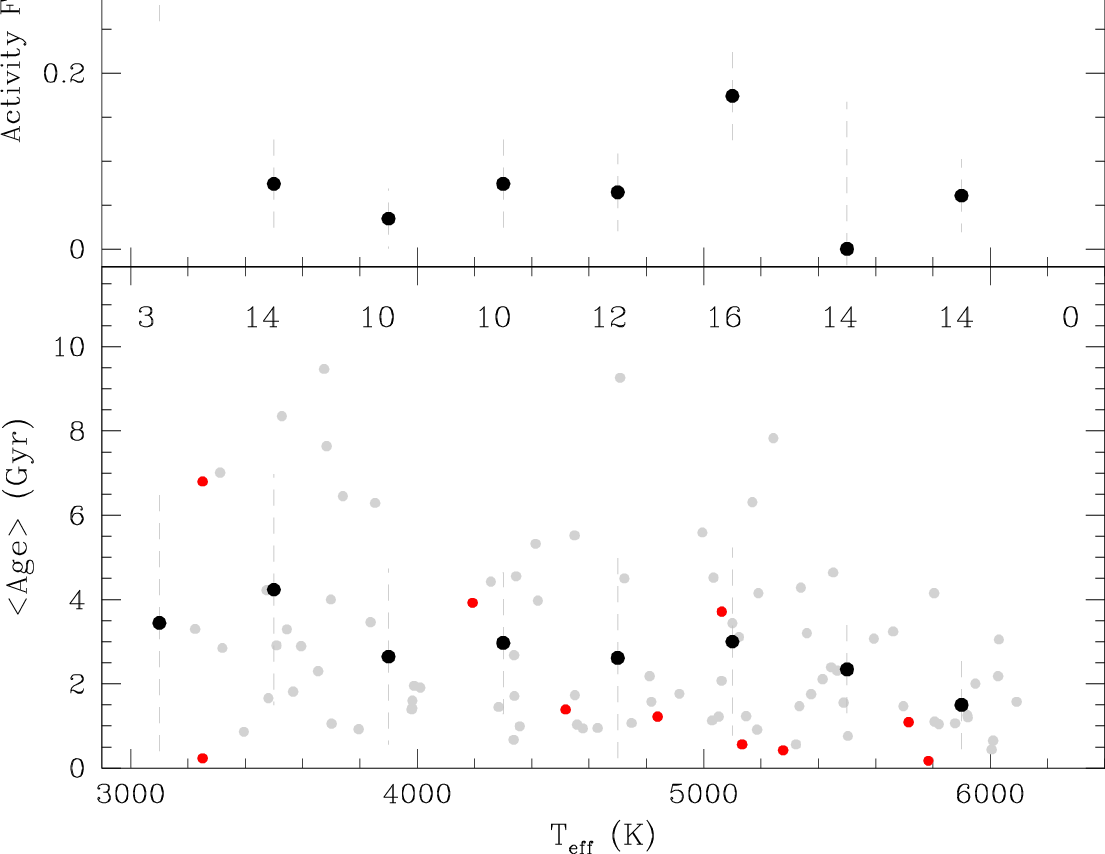}
    \caption{Top panels: activity fractions as a function of effective
      temperatures for  the entire sample of  main sequence companions
      to white dwarfs studied in this  work (left panel) and for those
      stars  that  we   observed  with  high-resolution  spectrographs
      (right). The  number of stars in  each bin are indicated  on the
      top of the  panels. Bottom panels: the distribution  of ages per
      temperature  bin (grey  dots;  red dots  indicate active  stars)
      together with the associated mean and standard deviations (black
      solid dots and lines, respectively). The number of stars per age
      bin are also indicated in the  top. Note that not all stars with
      derived effective temperatures have measured ages.}
    \label{fig:age-act2}
\end{figure*}

\section{Fraction of active stars}
\label{s-fractions}

We represent  the fraction  of active  stars as a  function of  age in
Figure\,\ref{fig:age-act1}.   The   top-left  panel   illustrates  the
fractions  for the  whole sample,  i.e.  stars  with activity  indexes
available and with ages associated  to uncertainties under 1\,Gyr. The
top-right panel shows the same  but considering only objects that were
observed  with high  resolution spectrographs,  which allow  detecting
weaker emission lines.  The comparison  reveals that, as expected, the
activity fractions  overall increase when we  consider high-resolution
spectra, but the pattern remains  the same. In particular, the highest
activity  fraction  corresponds  to  the  younger  bin  and  no  stars
displaying H$\alpha$  nor \Ion{Ca}{ii}~H\&K emission in  their spectra
are found older than 5\,Gyr. The exception is J1634+5710, which turned
out  to  be the  well-known  eclipsing  main sequence  binary  CM\,Dra
(implying this is a hierarchical triple system formed by an inner main
sequence binary and an outer DQ white dwarf companion) with an orbital
period of 1.27 days \citep{Lacy1977, Metcalfe1996, Morales2009} and an
age of 8.5$\pm$3.5\,Gyr \citep{Feiden2014},  in agreement with our age
derivation of 6.8$\pm$0.4\,Gyr from the DQ white dwarf.  In this case,
activity is  enhanced due to  the fast rotation  of the M  star rather
than by it being young.  The lack of systems displaying emission lines
above 5\,Gyr is consistent with an upper limit to the probability that
a star  is active  of 2  per cent, within  1$\sigma$. Even  though the
number of such old stars is only 25, this is a confirmation that young
stars are generally more active  presumably because they rotate faster
and that relatively old stars do  not appear to be magnetically active
since they  have been  further exposed to  magnetic braking  (with the
exception of stars that are  members of close binaries).  As expected,
the  activity fractions  gradually decrease  until 3\,Gyr,  however at
this point  they increase back,  which is at  odds with the  idea that
stars  become less  active  as  they grow  old.  This  feature can  be
understood as follows.

Given that  the lifetime of a  main sequence star depends  on its mass
(or effective  temperature), with increasing lifetimes  for decreasing
masses \citep[][see  their equations  A.10]{Tuffs2004}, the  spread of
ages  at  a  given  effective temperature  increases  with  decreasing
effective temperature.  Thus,  between 0.1 and 5\,Gyr our  sample is a
mix of  stars at  different effective  temperatures, $T_\mathrm{eff}$,
but    with    a    clear     concentration    of    hotter    objects
($T_\mathrm{eff}\ga$5000\,K)  at younger  ages  ($\la$2\,Gyr; see  the
bottom  panels  of  Figure\,\ref{fig:age-act1}).  Since  the  activity
lifetimes are also a function of mass, with massive (and hotter) stars
having  considerably  lower  activity  lifetimes  \citep{westetal08-1,
  reiners+mohanty12-1}, the  activity fractions abruptly  decrease for
ages in  the range  $\simeq$0.1--3\,Gyr.  The  fact that  hotter stars
($T_\mathrm{eff}\ga$5000\,K) are less  common after $\simeq$4--5\,Gyr,
combined with the  longer activity lifetimes of  cooler stars, implies
that the activity fractions increase back between $\simeq$3--5\,Gyr.

To further investigate the  relation between effective temperature and
activity we show in  Figure\,\ref{fig:age-act2} the activity fractions
as a  function of effective  temperature. The top-left  panel displays
the  fractions  for the  whole  sample  (stars with  activity  indexes
available and  with effective temperature uncertainties  under 120\,K)
and  the  top-right panel  the  fraction  for  those stars  that  were
observed with higher-resolution spectrographs.  The values are similar
except for  the coolest bin,  where the  percentage of stars  that are
active increases  considerably from  $\simeq$0.1 to  $\simeq$0.5. Even
though the  number of stars  associated to high-resolution  spectra at
this  bin  ($\simeq3000$\,K)  is  very   low,  this  increase  is  not
unexpected   since  we   are  near   the  fully   convective  boundary
\citep{Baraffe2018},  and  fully  convective stars  have  considerably
higher   activity  lifetimes   \citep{rebassa-mansergasetal13-1}.  The
fluctuations in the activity fractions  between 3000 and 6000\,K are a
consequence of the stars having different combinations of ages in each
bin (see  the bottom  panels of  Figure\,\ref{fig:age-act2}).

\section{The rotation-activity relation}
\label{s-actrot}

In this section we first  study the rotation-activity relation for our
sample  of   stars  with  measured  projected   rotational  velocities
(V$_\mathrm{rot}\,\sin i$) and H$\alpha$  equivalent widths.  Then, we
study  the same  relation  for those  stars  with measured  rotational
velocities and available S$_\mathrm{HK}$ indexes.

The strength of  magnetic activity was quantified as  the logarithm of
the  ratio  between  the   H$\alpha$  luminosity  and  the  bolometric
luminosity,  i.e.   $\log (L_\mathrm{H\alpha}/L_\mathrm{bol}$),  which
was obtained as
\begin{equation}
    L_\mathrm{H\alpha}/L_\mathrm{bol} = \chi \times |\mathrm{EW}_\mathrm{H\alpha}|, 
\end{equation}
\noindent
where $\mathrm{EW}_\mathrm{H\alpha}$ is the H$\alpha$ equivalent width
and $\chi$ is the ratio between  the continuum flux near the H$\alpha$
line and the bolometric flux at a given effective temperature. In this
exercise    we    adopted    the    $\chi$    values    provided    by
\citet{Fangetal2018}, which are defined for  a wide range of effective
temperatures  (from  2800  to   6800\,K).   Moreover,  the  rotational
velocities considered had  errors of less than 1\,\kms\  and the $\log
(L_\mathrm{H\alpha}/L_\mathrm{bol})$ values were  restricted to errors
no larger  than 0.75. It is  important to emphasise that,  previous to
the   mentioned   calculations,   the   $\mathrm{EW}_\mathrm{H\alpha}$
measurements  were  corrected  from  basal values  of  inactive  stars
following \citet{Fangetal2016}. In particular, we fitted a third order
polynomial to the $\mathrm{EW}_\mathrm{H\alpha}$  of inactive stars in
our sample  and the results from  the fit were subtracted  so that the
corrected  $\mathrm{EW}_\mathrm{H\alpha}$ values  were either  zero or
lower than that.

The   results  obtained   are  illustrated   in  the   top  panel   of
Figure\,\ref{fig:act-rot}          and          encompass          the
$L_\mathrm{H\alpha}/L_\mathrm{bol}$  versus rotation  relation for  43
FGK stars.  For comparison we also include in this figure the relation
we obtained in  Paper\,I for 47 M dwarfs in  tight binaries with white
dwarfs (blue  solid dots; note that  in this case the  orbital periods
were known and,  as a consequence, the rotational  velocities were not
affected   by   the   $\sin   i$    factor)   and   the   results   by
\citet{Reiners2012} for  242 single  M dwarfs with  measured projected
rotational velocities.

\begin{figure}
    \centering
    \includegraphics[angle=-90,width=\columnwidth]{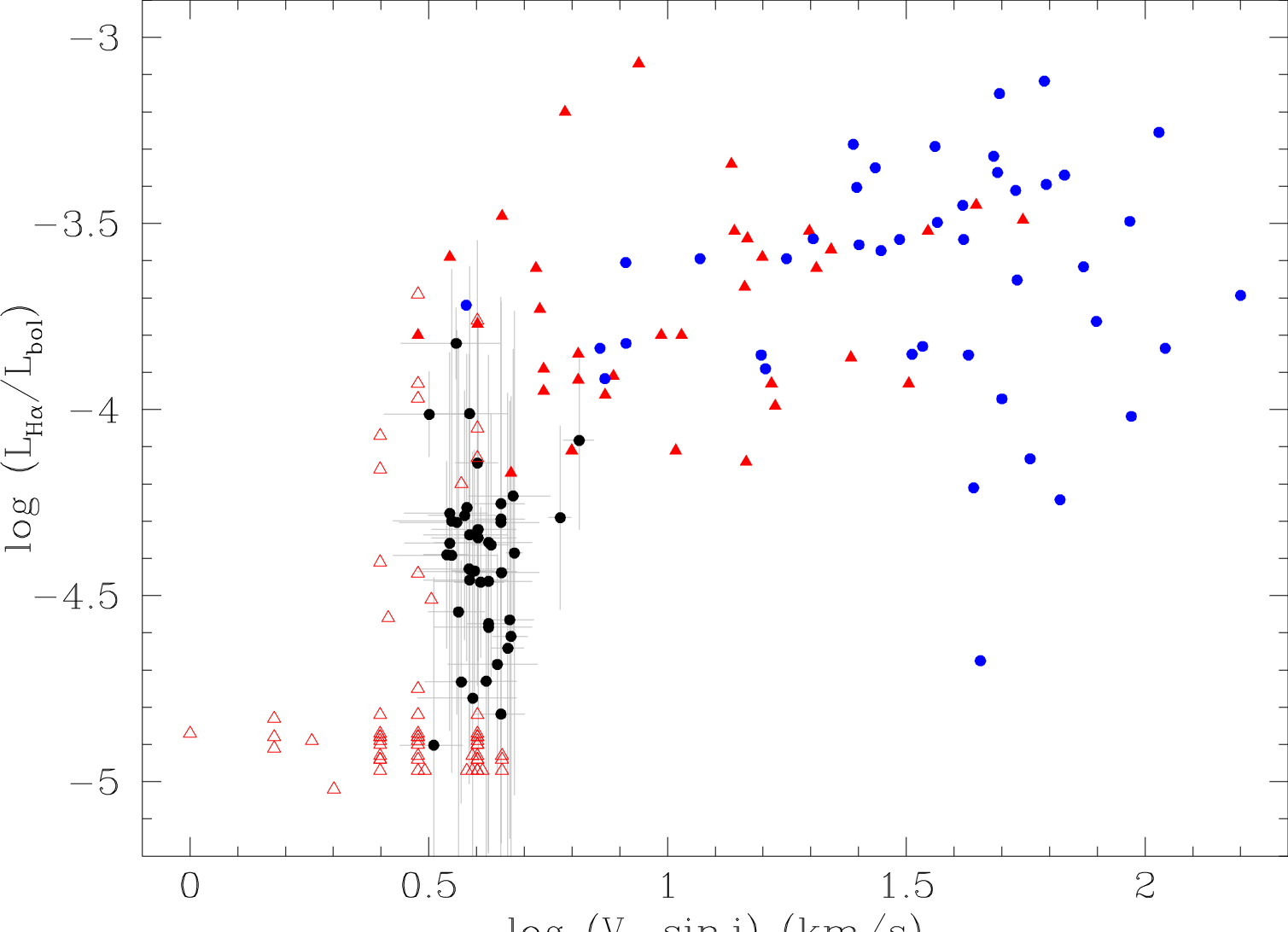}
    \includegraphics[angle=-90,width=\columnwidth]{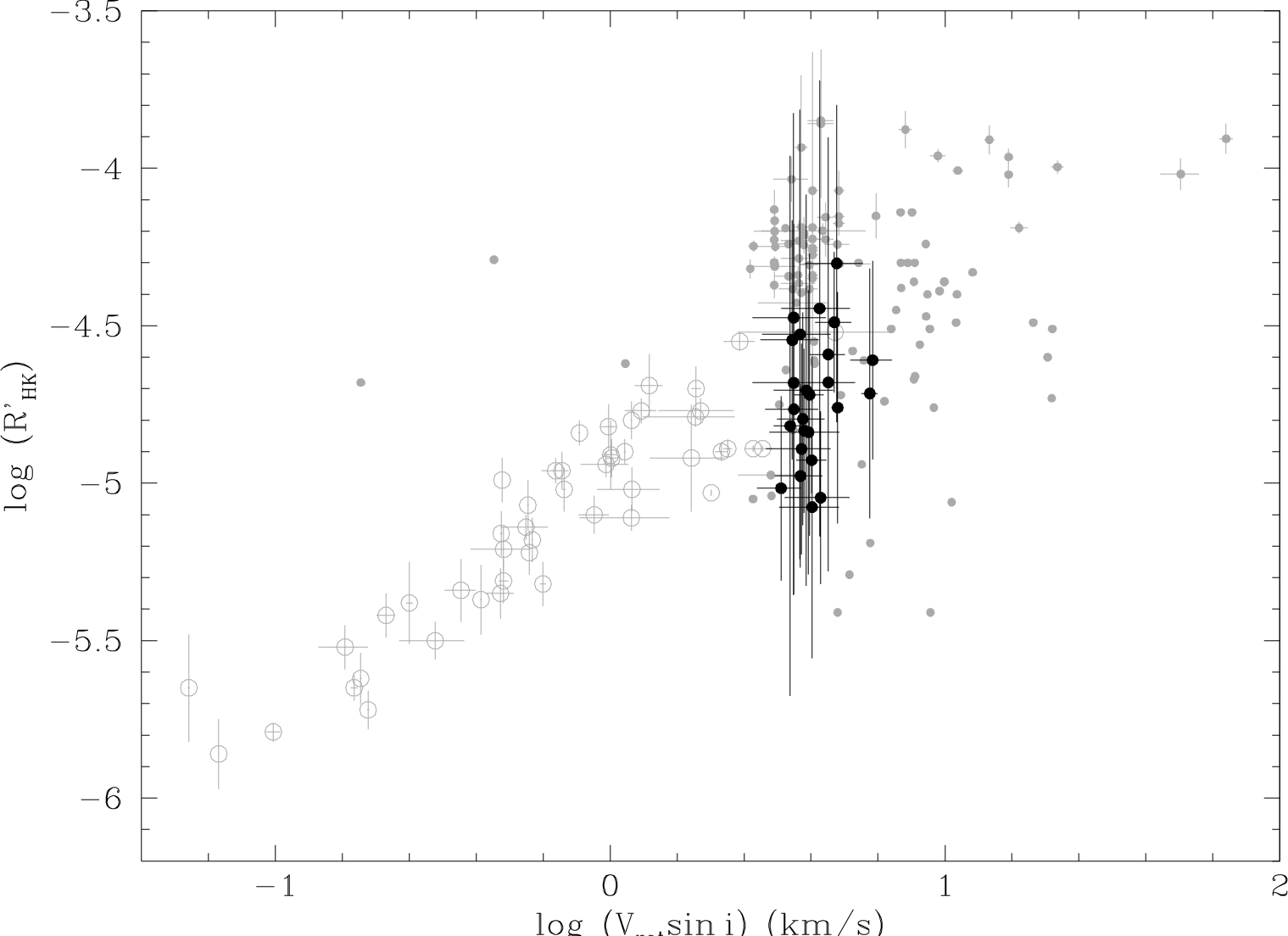}
    \caption{Top       panel:      $L_\mathrm{H\alpha}/L_\mathrm{bol}$
      (indicating the strength of magnetic  activity) as a function of
      projected rotation velocities for the 43 FGK stars considered in
      this work (black solid dots), the  47 M dwarfs in tight binaries
      with white dwarfs  from Paper\,I (blue solid dots;  note that in
      this cases  the rotational  velocities are  not affected  by the
      $\sin   i$  factor),   and  the   242  single   M  dwarfs   from
      \citet{Reiners2012}  (solid  red triangles  indicate  accurately
      measured velocities,  whilst open  red triangles  indicate upper
      limits).  Bottom  panel: log(R'$_\mathrm{HK}$) as a  function of
      projected rotation velocities for 70  stars in our sample (black
      solid dots).  Gray solid dots are data from \citet{Martinez2010}
      (no errors  provided since the  authors do not report  them) and
      \citet{Maldonado22}.    Solid  open   circles   are  data   from
      \citet{Suarez2015}  and  represent   rotational  velocities  not
      affected by the inclination factor.}
    \label{fig:act-rot}
\end{figure}

Whilst the  sample studied in  Paper\,I populates the regime  in which
the      strength      of       magnetic      activity      saturates,
i.e. log($L_\mathrm{H\alpha}/L_\mathrm{bol})  \sim -3.5$ independently
of rotation,  most of the stars  analysed in this work  are located in
the   area   expected   for    inactive   or   weakly   active   stars
($L_\mathrm{H\alpha}/L_\mathrm{bol}   <   -4$),  typical   of   slower
rotators.   Indeed, none  of the  43  FGK stars  display H$\alpha$  in
emission.   Given that  the full  sample of  stars considered  in this
exercise have FGK  spectral types, it is not  entirely surprising that
in some cases the strength  of magnetic activity reached values higher
than  $-$4. In  FGK stars  the H$\alpha$  line is  generally found  in
absorption and the depth of the line decreases as soon as the level of
magnetic activity  increases. In  other words,  the H$\alpha$  line is
radiation-dominated, i.e. the increase  of the optical depth initially
leads to  a deeper absorption  profile, until the electron  density is
high  enough  to  take   the  line  into  the  collisionally-dominated
formation   regime\footnote{This   is   always  the   case   for   the
  \Ion{Ca}{ii}~H\&K  lines,  in  which   the  radiated  flux  steadily
  increases with pressure.}. Only in those cases in which the level of
magnetic activity is very high the line will be seen in emission. As a
consequence, weakly  active FGK stars  will not display  emission, but
less absorption than inactive stars.  This implies that after applying
the basal  correction to the  measured $\mathrm{EW}_\mathrm{H\alpha}$,
the  corrected values  will  be  negative for  those  that are  weakly
active.   Thus,  the more  active  the  star,  the more  negative  the
corrected   $\mathrm{EW}_\mathrm{H\alpha}$   and    the   higher   the
$L_\mathrm{H\alpha}/L_\mathrm{bol}$  value.   We   conclude  that  the
results provided  by Paper\,I and  this work clearly support  the idea
that  the  strength  of  magnetic  activity  gradually  increases  for
increasing  rotation  and  that   it  saturates  when  the  rotational
velocities of the stars are higher than $\simeq$10\,\kms.

To quantify  the level of chromospheric  \Ion{Ca}{ii}~H\&K emission we
derived the  R'$_\mathrm{HK}$ index as introduced  by \citet[][see the
  Appendix]{noyesetal84-1},       which       is      defined       as
R'$_\mathrm{HK}=$R$_\mathrm{HK}$-R$_\mathrm{phot}$      (see      also
\citealt{Hall2007, Lovis2011,  Mittag2013}). R$_\mathrm{phot}$  is the
photospheric contribution,  which depends on  the (B-V) colour  of the
star.  R$_\mathrm{HK}$ gives the emission normalised to the bolometric
brightness of the star and it also depends on its (B-V) colour as well
as on the S$_\mathrm{HK}$ index on  the Mount Wilson scale.  Since the
S$_\mathrm{HK}$  values we  obtained in  Section\,\ref{sec:indicators}
are  not  in the  required  scale  and no  stars  in  our sample  have
available  Mount  Wilson  S$_\mathrm{HK}$ indexes  in  the  literature
\citep[e.g.][]{noyesetal84-1, Duncan1991}, we rederived them following
the procedure outlined in  \cite{Maldonado22} for TNG/HARPS-N spectra.
That is, we first reobtained the S$_\mathrm{HK}$ index for our targets
with  HARPS-N spectra  and converted  them to  the Mount  Wilson scale
using  equation  3 of  \cite{Maldonado22}.   During  this process,  we
obtained the (B-V) colours of  the stars interpolating their effective
temperatures in the updated  tables of \citet{Pecaut2013}, which gives
very similar results as using the (B-V)-effective temperature relation
provided  by \cite{Maldonado22}  in their  Appendix A.   Given that  9
objects in  our sample were  observed by the  TNG and LAMOST  with its
low-resolution  spectrograph,  we  obtained  a  relation  between  the
S$_\mathrm{HK}$ indexes obtained  in Section\,\ref{sec:indicators} and
the  corresponding  Mount  Wilson   S$_\mathrm{HK}$  values  (see  the
Appendix)\footnote{Only  three  objects with  S$_\mathrm{HK}$  HARPS-N
  values had  common observations with Mercator/HERMES,  and none with
  Xinglong   or  LAMOST   medium   resolution   spectrograph.   As   a
  consequence, we  could not  derive R'$_\mathrm{HK}$ for  those stars
  observed  with Mercator,  Xinglong  or  LAMOST medium  resolution.}.
Using this relation we managed  to derive R'$_\mathrm{HK}$ indexes for
all stars observed by the  low-resolution spectrograph of LAMOST, i.e.
a total of 150 stars (27 observed by the TNG and 123 by LAMOST).

The logarithm  of the  R'$_\mathrm{HK}$ values  above obtained  for 70
stars  (we do  not apply  any  restriction regarding  the errors)  are
represented  as a  function of  their available  rotational velocities
(only values with errors  of less than 1 km/s) in  the bottom panel of
Figure\,\ref{fig:act-rot}.   For  comparison,   we  also  include  the
measurements    by   \citet{Martinez2010},    \citet{Suarez2015}   and
\citet{Maldonado22}.   Note  that  \citet{Suarez2015} do  not  provide
rotational velocities  but rotational periods (P),  which we converted
into velocities via  2$\pi\times$R/P, where we adopted  the radius (R)
as  given  by  the  updated   tables  of  \citet{Pecaut2013}  for  the
corresponding available  spectral types.  As  it can be seen  from the
Figure, our R'$_\mathrm{HK}$  values fall where they  are expected for
their   rotational    velocities.    In   the   same    way   as   for
$L_\mathrm{H\alpha}/L_\mathrm{bol}$,  we   observe  a   saturation  of
log(R'$_\mathrm{HK}$) for fast rotators.

\section{The age-activity relation}

The strength of magnetic  activity -- again represented quantitatively
by log($L_\mathrm{H\alpha}/L_\mathrm{bol}$)  and log(R'$_\mathrm{HK}$)
-- is illustrated  as a function of  age in Figure\,\ref{fig:act-age}.
In the top panel of the  figure we considered main sequence stars with
errors in log($L_\mathrm{H\alpha}/L_\mathrm{bol}$)  under 0.75 and age
errors of  less than  1\,Gyr.  This  resulted in 87  FGK stars  and 37
partially convective M  dwarfs.  In the middle panel we  show 43 stars
(37 FGK  and 6 partially convective  M stars) with age  errors of less
than 1\,Gyr and  no error cuts applied  to their log(R'$_\mathrm{HK}$)
measurements.

\begin{figure}
    \centering
    \includegraphics[angle=-90,width=\columnwidth]{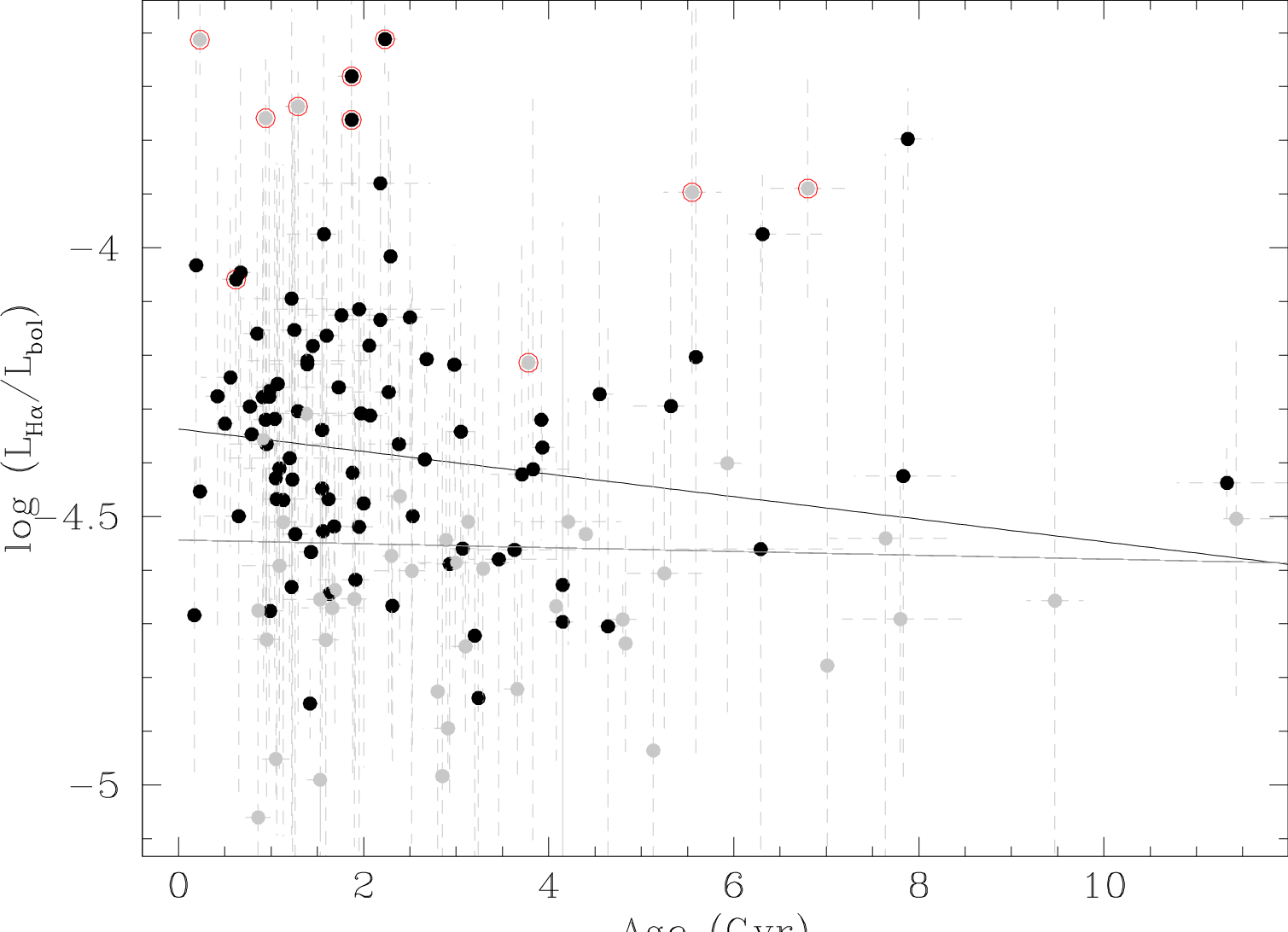}
    \includegraphics[angle=-90,width=\columnwidth]{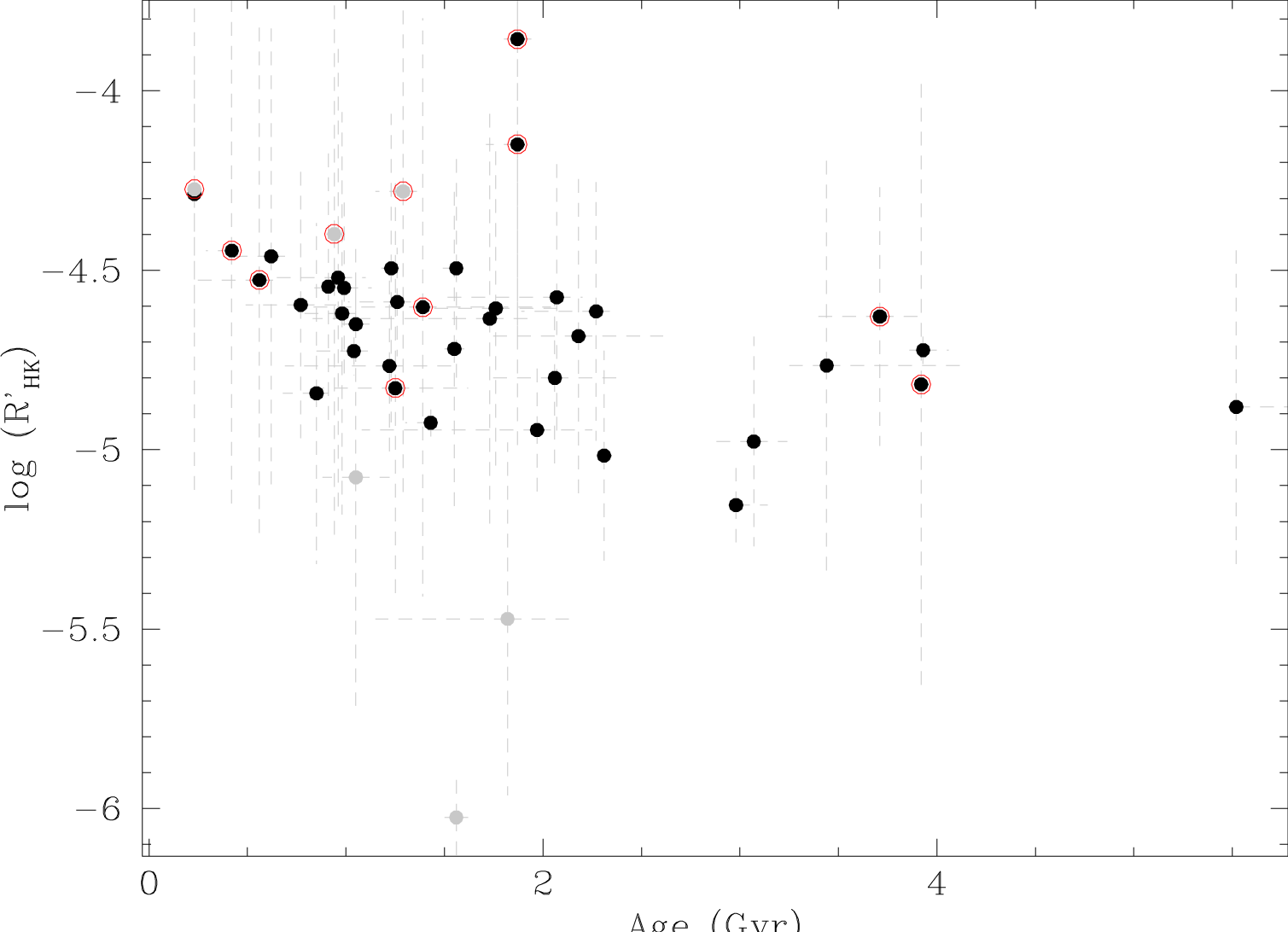}
    \includegraphics[angle=-90,width=\columnwidth]{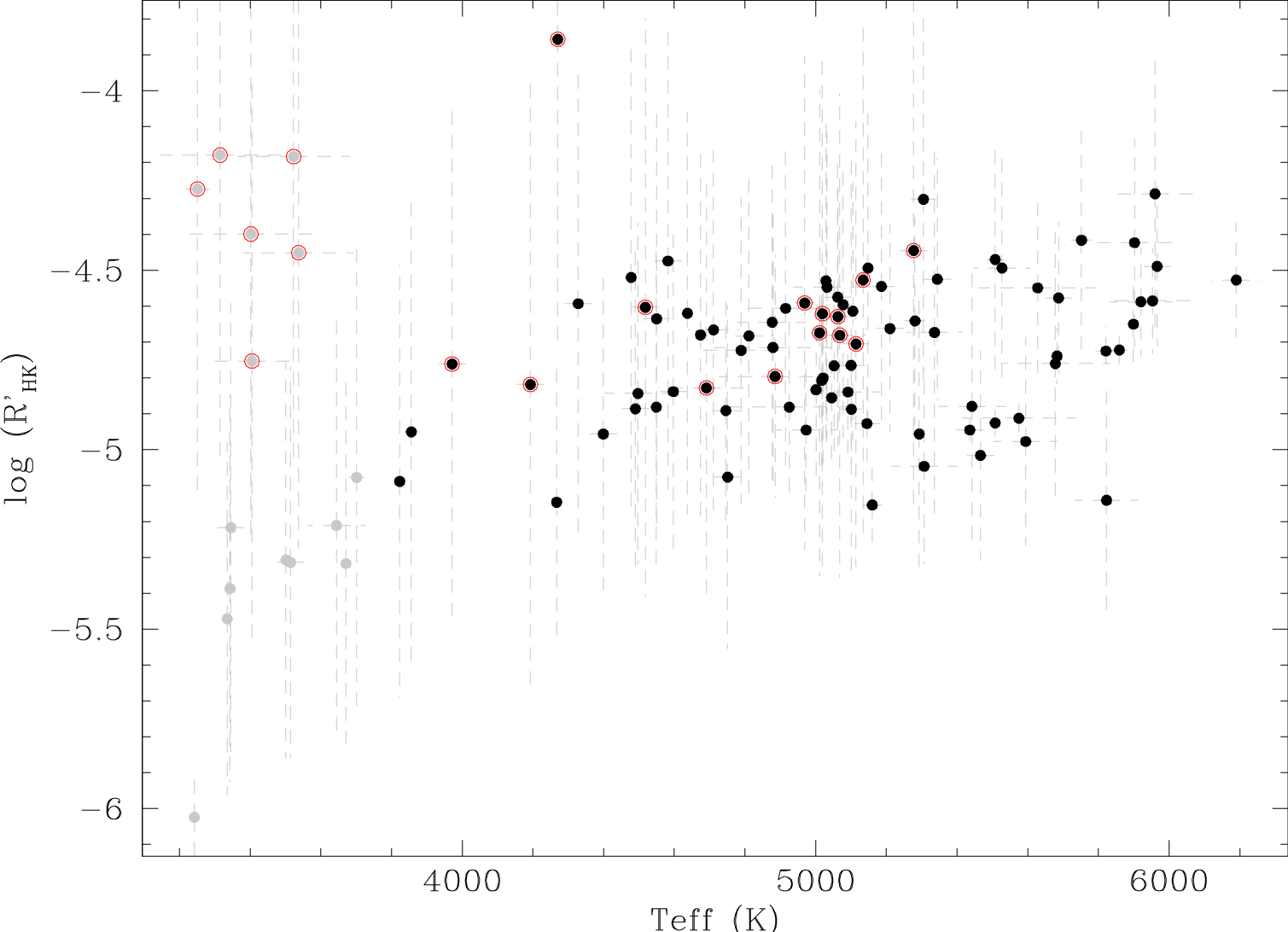}
    \caption{Top       panel:      $L_\mathrm{H\alpha}/L_\mathrm{bol}$
      (indicating the strength of magnetic  activity) as a function of
      age for  the 87 FGK  (black dots) and 37  partially-convective M
      stars (grey dots)  considered in this work. The  solid black and
      grey  lines  are a  linear  fit  to the  FGK  and  M star  data,
      respectively, which  illustrate the  tendency of M  stars having
      lower  values of  $L_\mathrm{H\alpha}/L_\mathrm{bol}$.  The  red
      open circles indicate  that H$\alpha$ in emission  is visible in
      the spectra.  Middle panel: log(R'$_\mathrm{HK}$) as  a function
      of  age for  37 FGK  (black dots)  and 6  partially-convective M
      stars  (grey  dots)  considered  in this  work.   Bottom  panel:
      log(R'$_\mathrm{HK}$) as a function of effective temperature for
      the  entire sample  of stars  studied this  work (restricted  to
      effective temperature errors of less than 200\,K).}
    \label{fig:act-age}
\end{figure}

Stars displaying H$\alpha$  in emission are shown as red  open dots in
the top panel  of Figure\,\ref{fig:act-age} and can mainly  be seen at
ages below  4\,Gyr.  The exceptions are  J0531+2142 (5.5$\pm$0.3\,Gyr)
and  J1634+5710 (6.8$\pm$0.4\,Gyr).   J1634+5710  is the  hierarchical
triple system composed of a  DQ plus the short-period eclipsing binary
CM\,Dra. As  mentioned earlier, in  this case  activity is due  to the
fast  rotation induced  by tidal  locking.  J0531+2142  has a  current
orbital projected separation of 1033 AU, therefore it is unlikely that
wind accretion  processes took place  in the  past. Given the  lack of
radial velocity measurements  due to the broad  H$\alpha$ emission, we
suggest this system  to be of similar nature as  J1634+5710.  The fact
that most active  stars are below 4\,Gyrs is in  line with our results
in Section\,\ref{s-fractions}, where we  found null activity fractions
for stars older than 5\,Gyr (with the obvious exception of J1634+5710;
see also Figure\,\ref{fig:age-act1}).  Moreover,  the vast majority of
the     stars      showing     H$\alpha$     in      emission     have
log($L_\mathrm{H\alpha}/L_\mathrm{bol})>-4$, as  expected. However, it
is  worth noting  that  4  FGK stars  with  similarly  high values  of
strength  of magnetic  activity do  not show  H$\alpha$ emission.   As
already  discussed  in the  previous  section,  this is  not  entirely
surprising  since  the  H$\alpha$  absorption  profile  in  FGK  stars
gradually decreases  for increasing  levels of magnetic  activity, and
only  for the  highest levels  can  it be  seen in  emission. This  is
generally not the case in M  dwarfs, which have much weaker absorption
profiles, and therefore are more prone to show emission once the level
of  magnetic   activity  increases.    Thus,  one  would   expect  the
log($L_\mathrm{H\alpha}/L_\mathrm{bol}$) values for  M stars to remain
roughly  similar for  inactive stars  and to  quickly rise  for active
stars,  and to  show  a wider  spread for  FGK  stars. Indeed,  visual
inspection of  Figure\,\ref{fig:act-age} shows a tendency  for M stars
to have lower log($L_\mathrm{H\alpha}/L_\mathrm{bol}$) values than FGK
stars. This is confirmed by the linear fits to the data represented as
black and grey solid lines for the FGK and M star data, respectively.

Further inspecting the top  panel of Figure\,\ref{fig:act-age} one can
see  that,   apart  from  J1634+5710   and  J0531+2142,  one   M  star
(J1043+5828)  with   an  age  of   $\simeq4$\,Gyr  and  3   FGK  stars
(J0707+2754, J0853+4406  and J1259+2528)  with ages  of $\simeq$2\,Gyr
display H$\alpha$  in emission.   The activity lifetimes  of partially
convective M stars are not expected  to be higher than 2\,Gyr and they
are  supposed to  be  even lower  for  FGK stars  \citep{westetal08-1,
  reiners+mohanty12-1}.   J1043+5828 has  a current  projected orbital
separation of 608  AU, therefore magnetic activity could  arise due to
faster rotation  induced by past  episodes of wind accretion  from the
white dwarf  precursor, when  the separation  was 2--5  times shorter.
Regarding the three FGK stars, their current projected separations are
4500 AU (J0707+2754),  6437 AU (J0853+4406) and  2935 AU (J1259+2528),
which indicates wind  accretion very likely did not take  place in the
past.  We argue  that the following physical mechanisms  may result in
these stars  rotating faster  than expected: (1)  it is  possible that
some of  them are suffering from  weakened magnetic braking and,  as a
consequence, are anomalously rapid rotators \citep{vanSaders2016}; (2)
nearby hidden unresolved companions may  exist that are massive enough
for the stars in the inner binary of these plausible triple systems to
be  partially  or  totally   tidally  locked,  thus  increasing  their
rotation.  Unfortunately, none of these stars have measured rotational
nor radial velocities (except J1259+2528 with just one radial velocity
value  of   $-19.20\pm$1.80  km/s)   to  confirm  or   disprove  these
hypotheses. It  is also plausible that  the ages are not  reliable for
J1043+5828, J0707+2754  and J1259+2528, since no  white dwarf spectral
types are available and the  evolutionary sequences employed might not
be adequate if they are not DAs.

The  middle   panel  of  Figure\,\ref{fig:act-age}  shows   a  similar
behaviour      of      log(R'$_\mathrm{HK}$)     with      age      as
log($L_\mathrm{H\alpha}/L_\mathrm{bol}$).       That      is,      the
log(R'$_\mathrm{HK}$)     values    are     generally    lower     for
partially-convective M stars than for FGK stars.  It is also important
to mention that active M stars have log(R'$_\mathrm{HK}$) measurements
considerably higher  than inactive ones,  whilst log(R'$_\mathrm{HK}$)
is similar  for both active and  inactive FGK stars.  This  is further
illustrated in the bottom panel of Figure\,\ref{fig:act-age}, where we
represent log(R'$_\mathrm{HK}$) as a function of effective temperature
for our entire  sample (restricted to effective  temperature errors of
less than 200\,K).

Finally, the  middle panel  of Figure\,\ref{fig:act-age}  also reveals
four   moderately   old   FGK    stars   (2--4\,Gyr)   which   display
\Ion{Ca}{ii}~H\&K emission  lines: J0221+5333,  J0707+2754, J0853+4406
and J1743+1252.  J0707+2754 and  J0853+4406 also display  H$\alpha$ in
emission and have been discussed above. J0221+5333 and J1743+1252 both
have two radial velocity measurements from spectra separated by 37 and
194 nights  respectively and  no variations  have been  detected.  The
current projected orbital separations are 333 AU (J0221+5333) and 1482
AU (J1743+1252), which indicates J0221+5333  might also be a candidate
for spin-up  rotation due to  past wind accretion  episodes.  However,
with a rotational  velocity of 3.45$\pm$0.37\,km/s this  does not seem
to  be the  case, unless  the measurement  is highly  affected by  the
inclination factor.

\section{The age-rotation relation}
\label{s-actrot}

We finally investigate the direct relation between age and rotation in
this  section.   We considered  all  stars  with projected  rotational
velocities  associated   to  errors   lower  than  1\,\kms\   and  age
uncertainties  under 1\,Gyr,  i.e.  38  FGK stars.   The relation  for
these objects  is illustrated  in Figure\,\ref{fig:age_rot}.   None of
them display H$\alpha$ nor \Ion{Ca}{ii}  in emission in their spectra,
which is in  line with the idea that slow  rotators like those studied
here  (V$_\mathrm{rot} \sin  i\la10$\,\kms)  are  inactive (or  weakly
active).  Even though there seems to  be a relatively large scatter of
velocities at  a given  age, there  is a  tendency for  decreasing the
rotational velocity with age (represented by  a linear fit to the data
in  Figure\,\ref{fig:age_rot}).  This  confirms the  expectations that
rotation  breaks   with  time,  i.e.   the   so-called  gyrochronology
\citep[e.g.][]{Fouque2023, Gaidos2023}, and at the same time indicates
that not  all stars may  suffer from weakened magnetic  braking, since
the oldest  in our sample ($>$6\,Gyr)  are indeed among the  ones with
the lowest velocities.

\section{Conclusions}

Binary stars composed of a white  dwarf and a main sequence companion,
WDMS binaries, are excellent tools for studying a wide variety of open
problems in  modern astronomy. In particular,  deriving reliable white
dwarf  ages  makes  them  ideal  objects  for  analysing  age  related
relations  such   as  the   age-metallicity  relation  in   the  solar
neighbourhood   \citep{Rebassa2016b,   rebassa-mansergasetal21},   the
age-velocity dispersion  relation \citep{Raddi2022} and, like  in this
work,  the age-activity-rotation  relation of  low-mass main  sequence
(FGKM) stars. Indeed, the results obtained  in this work from a sample
of WDMS  binaries in common  proper motion pairs identified  thanks to
the     $Gaia$      astrometry     and     in     our      Paper     I
\citep{rebassa-mansergasetal13-1} --  from a sample of  both close and
wide  binaries from  the Sloan  Digital  Sky Survey  -- have  provided
useful input for  improving our understanding between  the relation of
age, magnetism and rotation in low-mass stars.

\begin{figure}
    \centering
    \includegraphics[angle=-90,width=\columnwidth]{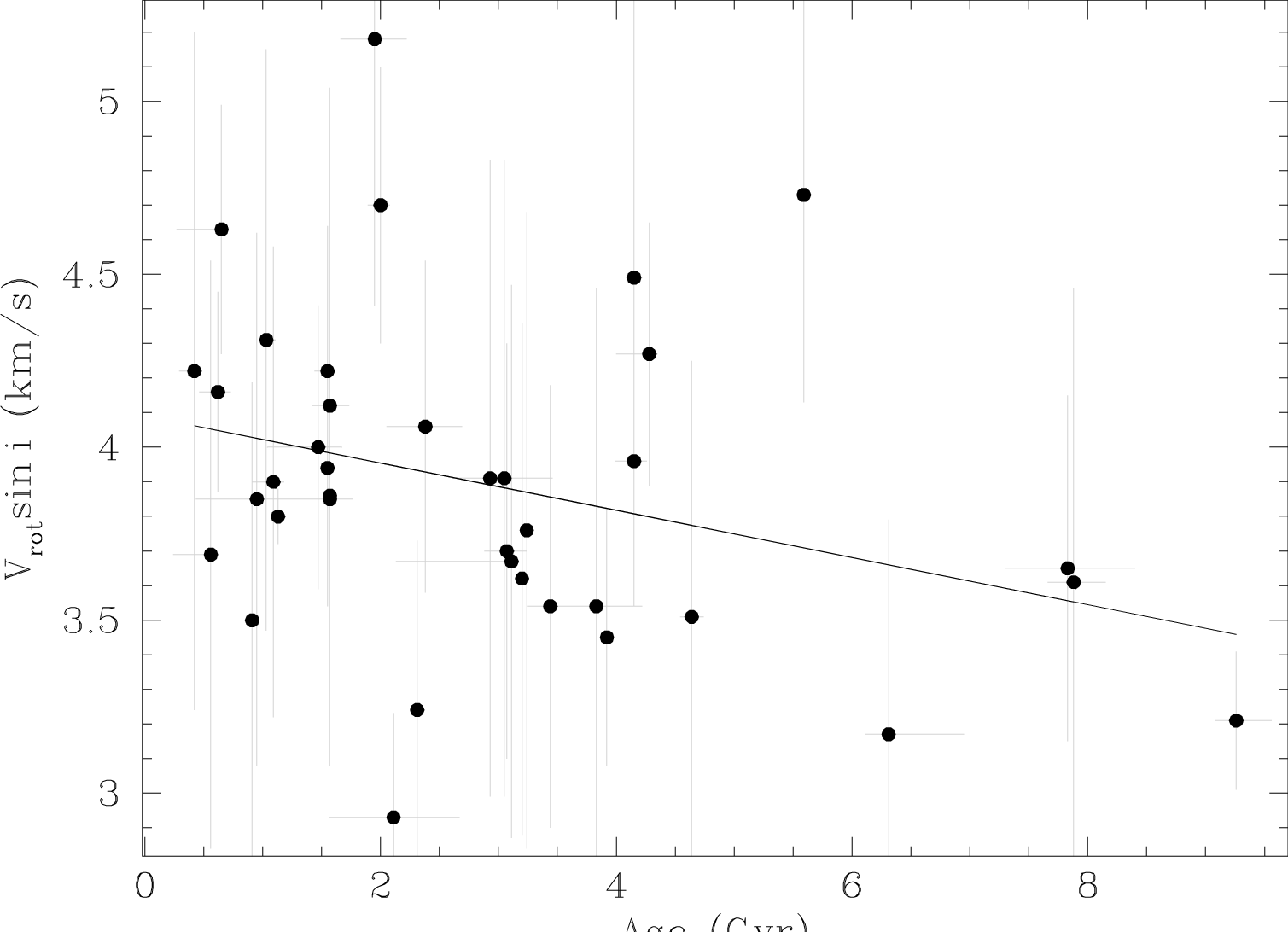}
    \caption{The projected  rotational velocity  as a function  of age
      for 38 FGK  stars. The solid black  line is a linear  fit to the
      data, which  indicates the tendency of  decreasing velocity with
      age.}
    \label{fig:age_rot}
\end{figure}

We have  shown that,  with the  exception of stars  that are  in close
binaries  (i.e.   hierarquical triple  systems  composed  of an  inner
binary plus an outer white dwarf), the activity fractions are null for
stars   older   than   5\,Gyr,  independently   of   their   effective
temperatures, and that the rotational velocities tend to decrease with
time.  This confirms the idea that  when stars grow old their level of
magnetic activity  decreases due  to the  effect of  magnetic braking,
which slows down their rotation.  We have also found that the activity
fractions   for   stars   younger  than   5\,Gyr   oscillate   between
$\simeq$10--40  per\,cent.   The  low   percentages  obtained  can  be
understood since the  stars in our sample are either  FGK or partially
convective  M  stars,  which  are  expected  to  have  short  activity
lifetimes. However, it  is worth noting that some stars  in our sample
are  active despite  being relatively  old, in  fact older  than their
expected activity lifetimes.  We argue  this indicates that a fraction
of stars  in our sample  might suffer from weakened  magnetic braking,
they may  had evolved through wind  accretion episodes in the  past or
they are associated to unreliable ages  due to the lack of white dwarf
spectral types.

We have also evaluated how  the strength of magnetic activity (studied
via   $L_\mathrm{H\alpha}/L_\mathrm{bol}$  and   the  R'$_\mathrm{HK}$
index) relates with both the  projected rotational velocities and age.
Given  that  our  stars   are  slow  rotators  (V$_\mathrm{rot}\sin  i
\la$10\,\kms)         their         measured         values         of
log($L_\mathrm{H\alpha}/L_\mathrm{bol}$)   and   log(R'$_\mathrm{HK}$)
fall  in  the   expected  region  for  inactive  M   stars  or  weakly
active/inactive        FGK       stars,        that       is        -4
$>$log($L_\mathrm{H\alpha}/L_\mathrm{bol}$),
(log(R'$_\mathrm{HK}$})$>$-5.

Although  the spread  of log($L_\mathrm{H\alpha}/L_\mathrm{bol}$)  and
log(R'$_\mathrm{HK}$) values  are nearly  constant with age,  they are
found  to be  higher for  FGK  than for  M  stars.  It  is also  worth
mentioning that,  as expected, the  vast majority of  stars displaying
H$\alpha$       in       emission        are       located       above
log$(L_\mathrm{H\alpha}/L_\mathrm{bol})=-4$.  Partially convective and
active M stars display  also considerably higher log(R'$_\mathrm{HK}$)
values than inactive ones. However, this is not generally the case for
FGK stars, in  which log(R'$_\mathrm{HK}$) is similar  for both active
and inactive stars.

Despite  the   fact  that   the  SDSS  WDMS   sample  we   studied  in
\citet{rebassa-mansergasetal13-1} and the  $Gaia$ common proper motion
pairs studied in  this work have been of  fundamental importance, they
come from very  different surveys and the  follow-up observations that
we have performed  were carried out using a wide  variety of different
telescopes and  spectroscopic resolutions.  As a  consequence, not all
our   targets  have   measured   H$\alpha$  fractional   luminosities,
R'$_\mathrm{HK}$    indexes,    rotational    velocities,    effective
temperatures  and  ages.   To  solve   this  issue,  the  4MOST  (4\,m
Multi-Object  Spectroscopic  Telescope) project  \citep{4MOST},  which
will start operations in early 2024,  will target, via its White Dwarf
Binary  Survey, $\simeq$2500  FGKM main  sequence companions  to white
dwarfs  in common  proper motion  pairs and  several thousand  WDMS in
tight orbits \citep{WDB2023}. Among the 2500 widely-separated binaries
that 4MOST will  observe, we expect $\simeq$10 per cent  to have total
ages approximately equal to their  cooling ages (i.e.  negligible main
sequence lifetimes).  This particular  sample will be extremely useful
in   mitigating   possible   age    uncertainties   related   to   the
initial-to-final mass relation.  Moreover, all the main sequence stars
will have available [Fe/H] values,  which will avoid age uncertainties
related to adopting a solar metallicity for the white dwarf precursors
in those  cases where the [Fe/H]  is unknown. The fact  that the white
dwarfs in  these common proper motion  pairs will also be  observed by
4MOST will allow us  to obtain a spectral type for  each of them, thus
avoiding unreliable ages due to wrong spectral type associations.  Not
only the  considerably higher  number of systems  to be  analysed, but
also the  observations to be carried  out in an homogeneous  way (thus
ensuring the  derivation of  the same parameters  for all  stars) will
allow us to investigate further and  in a more robust way the relation
between age, rotation and activity in low-mass main sequence stars.

\section*{Acknowledgments}

This work  has been  partially supported by  the Spanish  MINECO grant
PID2020-117252GB-I00 and  by the AGAUR/Generalitat de  Catalunya grant
SGR-386/2021.   RR acknowledges  support  from Grant  RYC2021-030837-I
funded  by  MCIN/AEI/  10.13039/501100011033 and  by  "European  Union
NextGenerationEU/PRTR".   MEC   acknowledges  grant  RYC2021-032721-I,
funded by  MCIN/AEI/10.13039/501100011033 and  by the  "European Union
NextGenerationEU/PRTR".  MAH was supported  by grant ST/V000853/1 from
the Science  and Technology  Facilities Council (STFC).   This project
has received  funding from the  European Research Council  (ERC) under
the European  Union's Horizon  2020 research and  innovation programme
(Grant agreement No. 101020057).

We thank the  anonymous referee for the suggestions  and comments that
helped improving the quality of the paper.

We  thank  Dr. Xiang-Song  Fang  for  sharing  his relation  of  basal
EW$_\mathrm{H\alpha}$  versus effective  temperature and  his relation
between effective temperature and $\chi_\mathrm{H\alpha}$.

Based on observations  made with the Telescopio  Nazionale Galileo and
Mercator  Telescope  awarded  to   the  International  Time  Programme
ITP18\_8. Based  on observations  performed with  the Xinglong  2.16 m
telescope.

Guoshoujing  Telescope   (the  Large   Sky  Area   Multi-Object  Fiber
Spectroscopic Telescope LAMOST) is a National Major Scientific Project
built by the Chinese Academy of  Sciences. Funding for the project has
been    provided   by    the   National    Development   and    Reform
Commission.   LAMOST  is   operated  and   managed  by   the  National
Astronomical Observatories, Chinese Academy of Sciences.

This work  has made use of  data from the European  Space Agency (ESA)
mission {\it  Gaia} (\url{https://www.cosmos.esa.int/gaia}), processed
by  the {\it  Gaia}  Data Processing  and  Analysis Consortium  (DPAC,
\url{https://www.cosmos.esa.int/web/gaia/dpac/consortium}).    Funding
for the DPAC has been provided by national institutions, in particular
the  institutions   participating  in  the  {\it   Gaia}  Multilateral
Agreement.

\section*{Data Availability}

The   data   underlying   this    article   are   available   in   the
manuscript.  Supplementary  material  will  be  shared  on  reasonable
request to the corresponding author.




\begin{thebibliography}{}
\makeatletter
\relax
\def\mn@urlcharsother{\let\do\@makeother \do\$\do\&\do\#\do\^\do\_\do\%\do\~}
\def\mn@doi{\begingroup\mn@urlcharsother \@ifnextchar [ {\mn@doi@}
  {\mn@doi@[]}}
\def\mn@doi@[#1]#2{\def\@tempa{#1}\ifx\@tempa\@empty \href
  {http://dx.doi.org/#2} {doi:#2}\else \href {http://dx.doi.org/#2} {#1}\fi
  \endgroup}
\def\mn@eprint#1#2{\mn@eprint@#1:#2::\@nil}
\def\mn@eprint@arXiv#1{\href {http://arxiv.org/abs/#1} {{\tt arXiv:#1}}}
\def\mn@eprint@dblp#1{\href {http://dblp.uni-trier.de/rec/bibtex/#1.xml}
  {dblp:#1}}
\def\mn@eprint@#1:#2:#3:#4\@nil{\def\@tempa {#1}\def\@tempb {#2}\def\@tempc
  {#3}\ifx \@tempc \@empty \let \@tempc \@tempb \let \@tempb \@tempa \fi \ifx
  \@tempb \@empty \def\@tempb {arXiv}\fi \@ifundefined
  {mn@eprint@\@tempb}{\@tempb:\@tempc}{\expandafter \expandafter \csname
  mn@eprint@\@tempb\endcsname \expandafter{\@tempc}}}

\bibitem[\protect\citeauthoryear{{Althaus}, {Camisassa}, {Miller Bertolami},
  {C{\'o}rsico}  \& {Garc{\'\i}a-Berro}}{{Althaus} et~al.}{2015}]{Althaus2015}
{Althaus} L.~G.,  {Camisassa} M.~E.,  {Miller Bertolami} M.~M.,  {C{\'o}rsico}
  A.~H.,   {Garc{\'\i}a-Berro} E.,  2015, \mn@doi [\aap]
  {10.1051/0004-6361/201424922}, \href
  {https://ui.adsabs.harvard.edu/abs/2015A&A...576A...9A} {576, A9}

\bibitem[\protect\citeauthoryear{{Anguiano}, {Freeman}, {Steinmetz}  \& {de
  Boer}}{{Anguiano} et~al.}{2010}]{anguianoetal10-1}
{Anguiano} B.,  {Freeman} K.~C.,  {Steinmetz} M.,   {de Boer} E.~W.,  2010, in
  {Block} D.~L.,  {Freeman} K.~C.,   {Puerari} I.,  eds, Galaxies and their
  Masks. p.~313, \mn@doi{$10.1007/978-1-4419-7317-7_26$}

\bibitem[\protect\citeauthoryear{{Angus}, {Aigrain}, {Foreman-Mackey}  \&
  {McQuillan}}{{Angus} et~al.}{2015}]{Angus2015}
{Angus} R.,  {Aigrain} S.,  {Foreman-Mackey} D.,   {McQuillan} A.,  2015,
  \mn@doi [\mnras] {10.1093/mnras/stv423}, \href
  {https://ui.adsabs.harvard.edu/abs/2015MNRAS.450.1787A} {450, 1787}

\bibitem[\protect\citeauthoryear{{Angus} et~al.,}{{Angus}
  et~al.}{2019}]{Angus2019}
{Angus} R.,  et~al., 2019, \mn@doi [\aj] {10.3847/1538-3881/ab3c53}, \href
  {https://ui.adsabs.harvard.edu/abs/2019AJ....158..173A} {158, 173}

\bibitem[\protect\citeauthoryear{{Angus} et~al.,}{{Angus}
  et~al.}{2020}]{Angus2020}
{Angus} R.,  et~al., 2020, \mn@doi [\aj] {10.3847/1538-3881/ab91b2}, \href
  {https://ui.adsabs.harvard.edu/abs/2020AJ....160...90A} {160, 90}

\bibitem[\protect\citeauthoryear{{Astudillo-Defru}, {Delfosse}, {Bonfils},
  {Forveille}, {Lovis}  \& {Rameau}}{{Astudillo-Defru}
  et~al.}{2017}]{Astudillo2017}
{Astudillo-Defru} N.,  {Delfosse} X.,  {Bonfils} X.,  {Forveille} T.,  {Lovis}
  C.,   {Rameau} J.,  2017, \mn@doi [\aap] {10.1051/0004-6361/201527078}, \href
  {https://ui.adsabs.harvard.edu/abs/2017A&A...600A..13A} {600, A13}

\bibitem[\protect\citeauthoryear{{Baliunas} et~al.,}{{Baliunas}
  et~al.}{1995}]{Baliunas1995}
{Baliunas} S.~L.,  et~al., 1995, \mn@doi [\apj] {10.1086/175072}, \href
  {https://ui.adsabs.harvard.edu/abs/1995ApJ...438..269B} {438, 269}

\bibitem[\protect\citeauthoryear{{Baraffe} \& {Chabrier}}{{Baraffe} \&
  {Chabrier}}{2018}]{Baraffe2018}
{Baraffe} I.,  {Chabrier} G.,  2018, \mn@doi [\aap]
  {10.1051/0004-6361/201834062}, \href
  {https://ui.adsabs.harvard.edu/abs/2018A&A...619A.177B} {619, A177}

\bibitem[\protect\citeauthoryear{{Barnes}}{{Barnes}}{2003}]{barnes03-1}
{Barnes} S.~A.,  2003, \mn@doi [\apjl] {10.1086/374681}, \href
  {http://adsabs.harvard.edu/abs/2003ApJ...586L.145B} {586, L145}

\bibitem[\protect\citeauthoryear{{Barnes}}{{Barnes}}{2010}]{Barnes2010}
{Barnes} S.~A.,  2010, \mn@doi [\apj] {10.1088/0004-637X/722/1/222}, \href
  {https://ui.adsabs.harvard.edu/abs/2010ApJ...722..222B} {722, 222}

\bibitem[\protect\citeauthoryear{{Barnes} \& {Kim}}{{Barnes} \&
  {Kim}}{2010}]{barnes+kim10-1}
{Barnes} S.~A.,  {Kim} Y.-C.,  2010, \mn@doi [\apj]
  {10.1088/0004-637X/721/1/675}, \href
  {http://adsabs.harvard.edu/abs/2010ApJ...721..675B} {721, 675}

\bibitem[\protect\citeauthoryear{{Barnes}, {Collier Cameron}, {Donati},
  {James}, {Marsden}  \& {Petit}}{{Barnes} et~al.}{2005}]{barnesetal05-1}
{Barnes} J.~R.,  {Collier Cameron} A.,  {Donati} J.-F.,  {James} D.~J.,
  {Marsden} S.~C.,   {Petit} P.,  2005, \mn@doi [\mnras]
  {10.1111/j.1745-3933.2005.08587.x}, \href
  {http://adsabs.harvard.edu/abs/2005MNRAS.357L...1B} {357, L1}

\bibitem[\protect\citeauthoryear{{Barrientos} \& {Chanam{\'e}}}{{Barrientos} \&
  {Chanam{\'e}}}{2021}]{Barrientos2021}
{Barrientos} M.,  {Chanam{\'e}} J.,  2021, arXiv e-prints, \href
  {https://ui.adsabs.harvard.edu/abs/2021arXiv210207790B} {p. arXiv:2102.07790}

\bibitem[\protect\citeauthoryear{{Boffin}}{{Boffin}}{2015}]{Boffin2015}
{Boffin} H. M.~J.,  2015, in {Boffin} H. M.~J.,  {Carraro} G.,   {Beccari} G.,
  eds,  Astrophysics and Space Science Library Vol. 413, Astrophysics and Space
  Science Library. p.~153 (\mn@eprint {arXiv} {1406.3473}),
  \mn@doi{10.1007/978-3-662-44434-4_7}

\bibitem[\protect\citeauthoryear{{Booth}, {Poppenhaeger}, {Watson}, {Silva
  Aguirre}, {Stello}  \& {Bruntt}}{{Booth} et~al.}{2020}]{Booth2020}
{Booth} R.~S.,  {Poppenhaeger} K.,  {Watson} C.~A.,  {Silva Aguirre} V.,
  {Stello} D.,   {Bruntt} H.,  2020, \mn@doi [\mnras] {10.1093/mnras/stz3039},
  \href {https://ui.adsabs.harvard.edu/abs/2020MNRAS.491..455B} {491, 455}

\bibitem[\protect\citeauthoryear{{Bowler} et~al.,}{{Bowler}
  et~al.}{2021}]{Bowler2021}
{Bowler} B.~P.,  et~al., 2021, \mn@doi [\aj] {10.3847/1538-3881/abd243}, \href
  {https://ui.adsabs.harvard.edu/abs/2021AJ....161..106B} {161, 106}

\bibitem[\protect\citeauthoryear{{Browning}}{{Browning}}{2008}]{browning08-1}
{Browning} M.~K.,  2008, \mn@doi [\apj] {10.1086/527432}, \href
  {http://adsabs.harvard.edu/abs/2008ApJ...676.1262B} {676, 1262}

\bibitem[\protect\citeauthoryear{{Browning}, {Basri}, {Marcy}, {West}  \&
  {Zhang}}{{Browning} et~al.}{2010}]{browningetal10-1}
{Browning} M.~K.,  {Basri} G.,  {Marcy} G.~W.,  {West} A.~A.,   {Zhang} J.,
  2010, \mn@doi [\aj] {10.1088/0004-6256/139/2/504}, \href
  {http://adsabs.harvard.edu/abs/2010AJ....139..504B} {139, 504}

\bibitem[\protect\citeauthoryear{{Camisassa}, {Althaus}, {C{\'o}rsico},
  {Vinyoles}, {Serenelli}, {Isern}, {Miller Bertolami}  \&
  {Garc{\'\i}a-Berro}}{{Camisassa} et~al.}{2016}]{Camisassa2016}
{Camisassa} M.~E.,  {Althaus} L.~G.,  {C{\'o}rsico} A.~H.,  {Vinyoles} N.,
  {Serenelli} A.~M.,  {Isern} J.,  {Miller Bertolami} M.~M.,
  {Garc{\'\i}a-Berro} E.,  2016, \mn@doi [\apj] {10.3847/0004-637X/823/2/158},
  \href {https://ui.adsabs.harvard.edu/abs/2016ApJ...823..158C} {823, 158}

\bibitem[\protect\citeauthoryear{{Camisassa}, {Althaus}, {Rohrmann},
  {Garc{\'\i}a-Berro}, {Torres}, {C{\'o}rsico}  \& {Wachlin}}{{Camisassa}
  et~al.}{2017}]{Camisassa2017}
{Camisassa} M.~E.,  {Althaus} L.~G.,  {Rohrmann} R.~D.,  {Garc{\'\i}a-Berro}
  E.,  {Torres} S.,  {C{\'o}rsico} A.~H.,   {Wachlin} F.~C.,  2017, \mn@doi
  [\apj] {10.3847/1538-4357/aa6797}, \href
  {https://ui.adsabs.harvard.edu/abs/2017ApJ...839...11C} {839, 11}

\bibitem[\protect\citeauthoryear{{Camisassa} et~al.,}{{Camisassa}
  et~al.}{2019}]{Camisassa2019}
{Camisassa} M.~E.,  et~al., 2019, \mn@doi [\aap] {10.1051/0004-6361/201833822},
  \href {https://ui.adsabs.harvard.edu/abs/2019A&A...625A..87C} {625, A87}

\bibitem[\protect\citeauthoryear{{Camisassa}, {Torres}, {Hollands}, {Koester},
  {Raddi}, {Althaus}  \& {Rebassa-Mansergas}}{{Camisassa}
  et~al.}{2023}]{Camisassa2023}
{Camisassa} M.,  {Torres} S.,  {Hollands} M.,  {Koester} D.,  {Raddi} R.,
  {Althaus} L.~G.,   {Rebassa-Mansergas} A.,  2023, \mn@doi [\aap]
  {10.1051/0004-6361/202346628}, \href
  {https://ui.adsabs.harvard.edu/abs/2023A&A...674A.213C} {674, A213}

\bibitem[\protect\citeauthoryear{{Capitanio}, {Lallement}, {Vergely},
  {Elyajouri}  \& {Monreal-Ibero}}{{Capitanio} et~al.}{2017}]{Capitanio2017}
{Capitanio} L.,  {Lallement} R.,  {Vergely} J.~L.,  {Elyajouri} M.,
  {Monreal-Ibero} A.,  2017, \mn@doi [\aap] {10.1051/0004-6361/201730831},
  \href {https://ui.adsabs.harvard.edu/abs/2017A&A...606A..65C} {606, A65}

\bibitem[\protect\citeauthoryear{{Catal{\'a}n}, {Isern}, {Garc{\'\i}a-Berro}
  \& {Ribas}}{{Catal{\'a}n} et~al.}{2008}]{Catalan08}
{Catal{\'a}n} S.,  {Isern} J.,  {Garc{\'\i}a-Berro} E.,   {Ribas} I.,  2008,
  \mn@doi [\mnras] {10.1111/j.1365-2966.2008.13356.x}, \href
  {https://ui.adsabs.harvard.edu/abs/2008MNRAS.387.1693C} {387, 1693}

\bibitem[\protect\citeauthoryear{{Chabrier} \& {K{\"u}ker}}{{Chabrier} \&
  {K{\"u}ker}}{2006}]{chabrier+kueker06-1}
{Chabrier} G.,  {K{\"u}ker} M.,  2006, \mn@doi [\aap]
  {10.1051/0004-6361:20042475}, \href
  {http://adsabs.harvard.edu/abs/2006A\%26A...446.1027C} {446, 1027}

\bibitem[\protect\citeauthoryear{{Chaplin} et~al.,}{{Chaplin}
  et~al.}{2014}]{Chaplin2014}
{Chaplin} W.~J.,  et~al., 2014, \mn@doi [\apjs] {10.1088/0067-0049/210/1/1},
  \href {https://ui.adsabs.harvard.edu/abs/2014ApJS..210....1C} {210, 1}

\bibitem[\protect\citeauthoryear{{Charbonneau}}{{Charbonneau}}{2005}]{charbonneau05-1}
{Charbonneau} P.,  2005, Living Reviews in Solar Physics, \href
  {http://adsabs.harvard.edu/abs/2005LRSP....2....2C} {2, 2}

\bibitem[\protect\citeauthoryear{{Cui} et~al.,}{{Cui}
  et~al.}{2012}]{Cui2012RAA}
{Cui} X.-Q.,  et~al., 2012, \mn@doi [Research in Astronomy and Astrophysics]
  {10.1088/1674-4527/12/9/003}, \href
  {https://ui.adsabs.harvard.edu/abs/2012RAA....12.1197C} {12, 1197}

\bibitem[\protect\citeauthoryear{{Cummings}, {Kalirai}, {Tremblay},
  {Ramirez-Ruiz}  \& {Choi}}{{Cummings} et~al.}{2018}]{Cummings2018}
{Cummings} J.~D.,  {Kalirai} J.~S.,  {Tremblay} P.~E.,  {Ramirez-Ruiz} E.,
  {Choi} J.,  2018, \mn@doi [\apj] {10.3847/1538-4357/aadfd6}, \href
  {https://ui.adsabs.harvard.edu/abs/2018ApJ...866...21C} {866, 21}

\bibitem[\protect\citeauthoryear{{Donati} et~al.,}{{Donati}
  et~al.}{2008}]{donatietal08-1}
{Donati} J.-F.,  et~al., 2008, \mn@doi [\mnras]
  {10.1111/j.1365-2966.2008.13799.x}, \href
  {http://adsabs.harvard.edu/abs/2008MNRAS.390..545D} {390, 545}

\bibitem[\protect\citeauthoryear{{Dravins}, {Lindegren}  \&
  {Torkelsson}}{{Dravins} et~al.}{1990}]{1990A&A...237..137D}
{Dravins} D.,  {Lindegren} L.,   {Torkelsson} U.,  1990, \aap, \href
  {https://ui.adsabs.harvard.edu/abs/1990A&A...237..137D} {237, 137}

\bibitem[\protect\citeauthoryear{{Dufour}, {Blouin}, {Coutu},
  {Fortin-Archambault}, {Thibeault}, {Bergeron}  \& {Fontaine}}{{Dufour}
  et~al.}{2017}]{Dufour2017}
{Dufour} P.,  {Blouin} S.,  {Coutu} S.,  {Fortin-Archambault} M.,  {Thibeault}
  C.,  {Bergeron} P.,   {Fontaine} G.,  2017, in {Tremblay} P.~E.,  {Gaensicke}
  B.,   {Marsh} T.,  eds,  Astronomical Society of the Pacific Conference
  Series Vol. 509, 20th European White Dwarf Workshop. p.~3 (\mn@eprint {arXiv}
  {1610.00986}), \mn@doi{10.48550/arXiv.1610.00986}

\bibitem[\protect\citeauthoryear{{Duncan} et~al.,}{{Duncan}
  et~al.}{1991}]{Duncan1991}
{Duncan} D.~K.,  et~al., 1991, \mn@doi [\apjs] {10.1086/191572}, \href
  {https://ui.adsabs.harvard.edu/abs/1991ApJS...76..383D} {76, 383}

\bibitem[\protect\citeauthoryear{{Durney} \& {Stenflo}}{{Durney} \&
  {Stenflo}}{1972}]{durney+stenflo72-1}
{Durney} B.~R.,  {Stenflo} J.~O.,  1972, \mn@doi [\apss] {10.1007/BF00649924},
  \href {http://adsabs.harvard.edu/abs/1972Ap%26SS..15..307D} {15, 307}

\bibitem[\protect\citeauthoryear{{Epstein} \& {Pinsonneault}}{{Epstein} \&
  {Pinsonneault}}{2014}]{Epstein2014}
{Epstein} C.~R.,  {Pinsonneault} M.~H.,  2014, \mn@doi [\apj]
  {10.1088/0004-637X/780/2/159}, \href
  {https://ui.adsabs.harvard.edu/abs/2014ApJ...780..159E} {780, 159}

\bibitem[\protect\citeauthoryear{{Fang}, {Zhao}, {Zhao}, {Chen}  \& {Bharat
  Kumar}}{{Fang} et~al.}{2016}]{Fangetal2016}
{Fang} X.-S.,  {Zhao} G.,  {Zhao} J.-K.,  {Chen} Y.-Q.,   {Bharat Kumar} Y.,
  2016, \mn@doi [\mnras] {10.1093/mnras/stw1923}, \href
  {https://ui.adsabs.harvard.edu/abs/2016MNRAS.463.2494F} {463, 2494}

\bibitem[\protect\citeauthoryear{{Fang}, {Zhao}, {Zhao}  \& {Bharat
  Kumar}}{{Fang} et~al.}{2018}]{Fangetal2018}
{Fang} X.-S.,  {Zhao} G.,  {Zhao} J.-K.,   {Bharat Kumar} Y.,  2018, \mn@doi
  [\mnras] {10.1093/mnras/sty212}, \href
  {https://ui.adsabs.harvard.edu/abs/2018MNRAS.476..908F} {476, 908}

\bibitem[\protect\citeauthoryear{{Feiden} \& {Chaboyer}}{{Feiden} \&
  {Chaboyer}}{2014}]{Feiden2014}
{Feiden} G.~A.,  {Chaboyer} B.,  2014, \mn@doi [\aap]
  {10.1051/0004-6361/201424288}, \href
  {https://ui.adsabs.harvard.edu/abs/2014A&A...571A..70F} {571, A70}

\bibitem[\protect\citeauthoryear{{Fouesneau}, {Rix}, {von Hippel}, {Hogg}  \&
  {Tian}}{{Fouesneau} et~al.}{2019}]{Fouesneau2019}
{Fouesneau} M.,  {Rix} H.-W.,  {von Hippel} T.,  {Hogg} D.~W.,   {Tian} H.,
  2019, \mn@doi [\apj] {10.3847/1538-4357/aaee74}, \href
  {https://ui.adsabs.harvard.edu/abs/2019ApJ...870....9F} {870, 9}

\bibitem[\protect\citeauthoryear{{Fouqu{\'e}} et~al.,}{{Fouqu{\'e}}
  et~al.}{2023}]{Fouque2023}
{Fouqu{\'e}} P.,  et~al., 2023, \mn@doi [\aap] {10.1051/0004-6361/202345839},
  \href {https://ui.adsabs.harvard.edu/abs/2023A&A...672A..52F} {672, A52}

\bibitem[\protect\citeauthoryear{{Gaia Collaboration} et~al.,}{{Gaia
  Collaboration} et~al.}{2018}]{Gaia2018}
{Gaia Collaboration} et~al., 2018, \mn@doi [\aap]
  {10.1051/0004-6361/201833051}, \href
  {https://ui.adsabs.harvard.edu/abs/2018A&A...616A...1G} {616, A1}

\bibitem[\protect\citeauthoryear{{Gaidos}, {Claytor}, {Dungee}, {Ali}  \&
  {Feiden}}{{Gaidos} et~al.}{2023}]{Gaidos2023}
{Gaidos} E.,  {Claytor} Z.,  {Dungee} R.,  {Ali} A.,   {Feiden} G.~A.,  2023,
  \mn@doi [\mnras] {10.1093/mnras/stad343}, \href
  {https://ui.adsabs.harvard.edu/abs/2023MNRAS.520.5283G} {520, 5283}

\bibitem[\protect\citeauthoryear{{Garc{\'\i}a-Berro}, {Torres}, {Isern},
  {Salaris}, {C{\'o}rsico}  \& {Althaus}}{{Garc{\'\i}a-Berro}
  et~al.}{2013}]{Garcia-Berro2013}
{Garc{\'\i}a-Berro} E.,  {Torres} S.,  {Isern} J.,  {Salaris} M.,
  {C{\'o}rsico} A.~H.,   {Althaus} L.~G.,  2013, in European Physical Journal
  Web of Conferences. p. 05003, \mn@doi{10.1051/epjconf/20134305003}

\bibitem[\protect\citeauthoryear{{Garc{\'\i}a-Zamora}, {Torres}  \&
  {Rebassa-Mansergas}}{{Garc{\'\i}a-Zamora} et~al.}{2023}]{Garcia-Zamora2023}
{Garc{\'\i}a-Zamora} E.~M.,  {Torres} S.,   {Rebassa-Mansergas} A.,  2023,
  arXiv e-prints, \href {https://ui.adsabs.harvard.edu/abs/2023arXiv230807090G}
  {p. arXiv:2308.07090}

\bibitem[\protect\citeauthoryear{{Garc{\'\i}a} et~al.,}{{Garc{\'\i}a}
  et~al.}{2014}]{Garcia2014}
{Garc{\'\i}a} R.~A.,  et~al., 2014, \mn@doi [\aap]
  {10.1051/0004-6361/201423888}, \href
  {https://ui.adsabs.harvard.edu/abs/2014A&A...572A..34G} {572, A34}

\bibitem[\protect\citeauthoryear{{Garraffo} et~al.,}{{Garraffo}
  et~al.}{2018}]{Garrafo2018}
{Garraffo} C.,  et~al., 2018, \mn@doi [\apj] {10.3847/1538-4357/aace5d}, \href
  {https://ui.adsabs.harvard.edu/abs/2018ApJ...862...90G} {862, 90}

\bibitem[\protect\citeauthoryear{{Gratton} et~al.,}{{Gratton}
  et~al.}{2021}]{Gratton2021}
{Gratton} R.,  et~al., 2021, \mn@doi [\aap] {10.1051/0004-6361/202039601},
  \href {https://ui.adsabs.harvard.edu/abs/2021A&A...646A..61G} {646, A61}

\bibitem[\protect\citeauthoryear{{Gray}}{{Gray}}{2008}]{2008oasp.book.....G}
{Gray} D.~F.,  2008, {The Observation and Analysis of Stellar Photospheres}

\bibitem[\protect\citeauthoryear{{Guti{\'e}rrez Albarr{\'a}n}
  et~al.,}{{Guti{\'e}rrez Albarr{\'a}n} et~al.}{2020}]{Gutierrez-Albarran2020}
{Guti{\'e}rrez Albarr{\'a}n} M.~L.,  et~al., 2020, \mn@doi [\aap]
  {10.1051/0004-6361/202037620}, \href
  {https://ui.adsabs.harvard.edu/abs/2020A&A...643A..71G} {643, A71}

\bibitem[\protect\citeauthoryear{{Hall}, {Lockwood}  \& {Skiff}}{{Hall}
  et~al.}{2007}]{Hall2007}
{Hall} J.~C.,  {Lockwood} G.~W.,   {Skiff} B.~A.,  2007, \mn@doi [\aj]
  {10.1086/510356}, \href
  {https://ui.adsabs.harvard.edu/abs/2007AJ....133..862H} {133, 862}

\bibitem[\protect\citeauthoryear{{Hall} et~al.,}{{Hall}
  et~al.}{2021}]{Hall2021}
{Hall} O.~J.,  et~al., 2021, arXiv e-prints, \href
  {https://ui.adsabs.harvard.edu/abs/2021arXiv210410919H} {p. arXiv:2104.10919}

\bibitem[\protect\citeauthoryear{{Hartmann} \& {Noyes}}{{Hartmann} \&
  {Noyes}}{1987}]{hartmann+noyes87-1}
{Hartmann} L.~W.,  {Noyes} R.~W.,  1987, \mn@doi [\araa]
  {10.1146/annurev.aa.25.090187.001415}, \href
  {http://adsabs.harvard.edu/abs/1987ARA\%26A..25..271H} {25, 271}

\bibitem[\protect\citeauthoryear{{Heintz}, {Hermes}, {El-Badry}, {Walsh}, {van
  Saders}, {Fields}  \& {Koester}}{{Heintz} et~al.}{2022}]{Heintz2022}
{Heintz} T.~M.,  {Hermes} J.~J.,  {El-Badry} K.,  {Walsh} C.,  {van Saders}
  J.~L.,  {Fields} C.~E.,   {Koester} D.,  2022, \mn@doi [\apj]
  {10.3847/1538-4357/ac78d9}, \href
  {https://ui.adsabs.harvard.edu/abs/2022ApJ...934..148H} {934, 148}

\bibitem[\protect\citeauthoryear{{Houdebine}, {Mullan}, {Bercu}, {Paletou}  \&
  {Gebran}}{{Houdebine} et~al.}{2017}]{Houdebine2017}
{Houdebine} E.~R.,  {Mullan} D.~J.,  {Bercu} B.,  {Paletou} F.,   {Gebran} M.,
  2017, \mn@doi [\apj] {10.3847/1538-4357/aa5cad}, \href
  {https://ui.adsabs.harvard.edu/abs/2017ApJ...837...96H} {837, 96}

\bibitem[\protect\citeauthoryear{{Jeans}}{{Jeans}}{1924}]{Jeans24}
{Jeans} J.~H.,  1924, \mn@doi [\mnras] {10.1093/mnras/85.1.2}, \href
  {https://ui.adsabs.harvard.edu/abs/1924MNRAS..85....2J} {85, 2}

\bibitem[\protect\citeauthoryear{{Jeffries} \& {Stevens}}{{Jeffries} \&
  {Stevens}}{1996}]{Jeffries1996}
{Jeffries} R.~D.,  {Stevens} I.~R.,  1996, \mn@doi [\mnras]
  {10.1093/mnras/279.1.180}, \href
  {https://ui.adsabs.harvard.edu/abs/1996MNRAS.279..180J} {279, 180}

\bibitem[\protect\citeauthoryear{{Jim{\'e}nez-Esteban}, {Torres},
  {Rebassa-Mansergas}, {Cruz}, {Murillo-Ojeda}, {Solano}, {Rodrigo}  \&
  {Camisassa}}{{Jim{\'e}nez-Esteban} et~al.}{2023}]{Jimenez2023}
{Jim{\'e}nez-Esteban} F.~M.,  {Torres} S.,  {Rebassa-Mansergas} A.,  {Cruz} P.,
   {Murillo-Ojeda} R.,  {Solano} E.,  {Rodrigo} C.,   {Camisassa} M.~E.,  2023,
  \mn@doi [\mnras] {10.1093/mnras/stac3382}, \href
  {https://ui.adsabs.harvard.edu/abs/2023MNRAS.518.5106J} {518, 5106}

\bibitem[\protect\citeauthoryear{{Kawaler}}{{Kawaler}}{1988}]{kawaler88-1}
{Kawaler} S.~D.,  1988, \mn@doi [\apj] {10.1086/166740}, \href
  {http://adsabs.harvard.edu/abs/1988ApJ...333..236K} {333, 236}

\bibitem[\protect\citeauthoryear{{Kiman} et~al.,}{{Kiman}
  et~al.}{2021}]{Kiman2021}
{Kiman} R.,  et~al., 2021, \mn@doi [\aj] {10.3847/1538-3881/abf561}, \href
  {https://ui.adsabs.harvard.edu/abs/2021AJ....161..277K} {161, 277}

\bibitem[\protect\citeauthoryear{{Kraft}}{{Kraft}}{1967a}]{kraft67-1}
{Kraft} R.~P.,  1967a, \mn@doi [\apj] {10.1086/149359}, \href
  {http://adsabs.harvard.edu/abs/1967ApJ...150..551K} {150, 551}

\bibitem[\protect\citeauthoryear{{Kraft}}{{Kraft}}{1967b}]{Kraft1967}
{Kraft} R.~P.,  1967b, \mn@doi [\apj] {10.1086/149359}, \href
  {http://adsabs.harvard.edu/abs/1967ApJ...150..551K} {150, 551}

\bibitem[\protect\citeauthoryear{{Lacy}}{{Lacy}}{1977}]{Lacy1977}
{Lacy} C.~H.,  1977, \mn@doi [\apj] {10.1086/155698}, \href
  {https://ui.adsabs.harvard.edu/abs/1977ApJ...218..444L} {218, 444}

\bibitem[\protect\citeauthoryear{{Lallement}, {Vergely}, {Valette},
  {Puspitarini}, {Eyer}  \& {Casagrande}}{{Lallement}
  et~al.}{2014}]{Lallement2014}
{Lallement} R.,  {Vergely} J.~L.,  {Valette} B.,  {Puspitarini} L.,  {Eyer} L.,
    {Casagrande} L.,  2014, \mn@doi [\aap] {10.1051/0004-6361/201322032}, \href
  {https://ui.adsabs.harvard.edu/abs/2014A&A...561A..91L} {561, A91}

\bibitem[\protect\citeauthoryear{{Lam}, {Hambly}, {Lodieu}, {Blouin}, {Harvey},
  {Smith}, {G{\'a}lvez-Ortiz}  \& {Zhang}}{{Lam} et~al.}{2020}]{Lam2020}
{Lam} M.~C.,  {Hambly} N.~C.,  {Lodieu} N.,  {Blouin} S.,  {Harvey} E.~J.,
  {Smith} R.~J.,  {G{\'a}lvez-Ortiz} M.~C.,   {Zhang} Z.~H.,  2020, \mn@doi
  [\mnras] {10.1093/mnras/staa584}, \href
  {https://ui.adsabs.harvard.edu/abs/2020MNRAS.493.6001L} {493, 6001}

\bibitem[\protect\citeauthoryear{{Leighton}}{{Leighton}}{1969}]{leighton69-1}
{Leighton} R.~B.,  1969, \mn@doi [\apj] {10.1086/149943}, \href
  {http://adsabs.harvard.edu/abs/1969ApJ...156....1L} {156, 1}

\bibitem[\protect\citeauthoryear{{Leiner}, {Mathieu}, {Gosnell}  \&
  {Sills}}{{Leiner} et~al.}{2018}]{Leiner2018}
{Leiner} E.,  {Mathieu} R.~D.,  {Gosnell} N.~M.,   {Sills} A.,  2018, \mn@doi
  [\apjl] {10.3847/2041-8213/aaf4ed}, \href
  {https://ui.adsabs.harvard.edu/abs/2018ApJ...869L..29L} {869, L29}

\bibitem[\protect\citeauthoryear{L{\'o}pez-Sanjuan et
    al.}{2022}]{Lopez2022} L{\'o}pez-Sanjuan C., Tremblay P.-E.,
  Ederoclite A., V{\'a}zquez Rami{\'o} H., Carrasco J.~M., Varela J.,
  Cenarro A.~J., et al., 2022, A\&A, 658,
  A79. doi:10.1051/0004-6361/202141746
  
\bibitem[\protect\citeauthoryear{{Lovis} et~al.,}{{Lovis}
  et~al.}{2011}]{Lovis2011}
{Lovis} C.,  et~al., 2011, \mn@doi [arXiv e-prints] {10.48550/arXiv.1107.5325},
  \href {https://ui.adsabs.harvard.edu/abs/2011arXiv1107.5325L} {p.
  arXiv:1107.5325}

\bibitem[\protect\citeauthoryear{McCleery et al.}{2020}]{McCleery2020}
  McCleery J., Tremblay P.-E., Gentile Fusillo N.~P., Hollands M.~A.,
  G{\"a}nsicke B.~T., Izquierdo P., Toonen S., et al., 2020, MNRAS,
  499, 1890. doi:10.1093/mnras/staa2030
  
\bibitem[\protect\citeauthoryear{{Magaudda}, {Stelzer}, {Covey}, {Raetz},
  {Matt}  \& {Scholz}}{{Magaudda} et~al.}{2020}]{Magaudda2020}
{Magaudda} E.,  {Stelzer} B.,  {Covey} K.~R.,  {Raetz} S.,  {Matt} S.~P.,
  {Scholz} A.,  2020, \mn@doi [\aap] {10.1051/0004-6361/201937408}, \href
  {https://ui.adsabs.harvard.edu/abs/2020A&A...638A..20M} {638, A20}

\bibitem[\protect\citeauthoryear{{Maldonado}, {Mart{\'\i}nez-Arn{\'a}iz},
  {Eiroa}, {Montes}  \& {Montesinos}}{{Maldonado} et~al.}{2010}]{Maldonado2010}
{Maldonado} J.,  {Mart{\'\i}nez-Arn{\'a}iz} R.~M.,  {Eiroa} C.,  {Montes} D.,
  {Montesinos} B.,  2010, \mn@doi [\aap] {10.1051/0004-6361/201014948}, \href
  {https://ui.adsabs.harvard.edu/abs/2010A&A...521A..12M} {521, A12}

\bibitem[\protect\citeauthoryear{{Maldonado} et~al.,}{{Maldonado}
  et~al.}{2022}]{Maldonado22}
{Maldonado} J.,  et~al., 2022, \mn@doi [\aap] {10.1051/0004-6361/202243360},
  \href {https://ui.adsabs.harvard.edu/abs/2022A&A...663A.142M} {663, A142}
  
\bibitem[\protect\citeauthoryear{{Mamajek} \& {Hillenbrand}}{{Mamajek} \&
  {Hillenbrand}}{2008}]{Mamajek+Hillenbrand2008}
{Mamajek} E.~E.,  {Hillenbrand} L.~A.,  2008, \mn@doi [\apj] {10.1086/591785},
  \href {https://ui.adsabs.harvard.edu/abs/2008ApJ...687.1264M} {687, 1264}

\bibitem[\protect\citeauthoryear{{Mart{\'\i}nez-Arn{\'a}iz}, {Maldonado},
  {Montes}, {Eiroa}  \& {Montesinos}}{{Mart{\'\i}nez-Arn{\'a}iz}
  et~al.}{2010}]{Martinez2010}
{Mart{\'\i}nez-Arn{\'a}iz} R.,  {Maldonado} J.,  {Montes} D.,  {Eiroa} C.,
  {Montesinos} B.,  2010, \mn@doi [\aap] {10.1051/0004-6361/200913725}, \href
  {https://ui.adsabs.harvard.edu/abs/2010A&A...520A..79M} {520, A79}

\bibitem[\protect\citeauthoryear{{Meibom}, {Barnes}, {Platais}, {Gilliland},
  {Latham}  \& {Mathieu}}{{Meibom} et~al.}{2015}]{Meibom2015}
{Meibom} S.,  {Barnes} S.~A.,  {Platais} I.,  {Gilliland} R.~L.,  {Latham}
  D.~W.,   {Mathieu} R.~D.,  2015, \mn@doi [\nat] {10.1038/nature14118}, \href
  {https://ui.adsabs.harvard.edu/abs/2015Natur.517..589M} {517, 589}

\bibitem[\protect\citeauthoryear{{Mestel}}{{Mestel}}{1984}]{mestel84-1}
{Mestel} L.,  1984, in {Baliunas} S.~L.,  {Hartmann} L.,  eds,  Lecture Notes
  in Physics, Berlin Springer Verlag Vol. 193, Cool Stars, Stellar Systems, and
  the Sun. p.~49, \mn@doi{10.1007/3-540-12907-3_179}

\bibitem[\protect\citeauthoryear{{Mestel} \& {Spruit}}{{Mestel} \&
  {Spruit}}{1987}]{mestel+spruit87-1}
{Mestel} L.,  {Spruit} H.~C.,  1987, 226, 57

\bibitem[\protect\citeauthoryear{{Metcalfe}, {Mathieu}, {Latham}  \&
  {Torres}}{{Metcalfe} et~al.}{1996}]{Metcalfe1996}
{Metcalfe} T.~S.,  {Mathieu} R.~D.,  {Latham} D.~W.,   {Torres} G.,  1996,
  \mn@doi [\apj] {10.1086/176657}, \href
  {https://ui.adsabs.harvard.edu/abs/1996ApJ...456..356M} {456, 356}

\bibitem[\protect\citeauthoryear{{Metcalfe} et~al.,}{{Metcalfe}
  et~al.}{2012}]{Metcalfe2012}
{Metcalfe} T.~S.,  et~al., 2012, \mn@doi [\apjl] {10.1088/2041-8205/748/1/L10},
  \href {https://ui.adsabs.harvard.edu/abs/2012ApJ...748L..10M} {748, L10}

\bibitem[\protect\citeauthoryear{{Metcalfe} et~al.,}{{Metcalfe}
  et~al.}{2020}]{Metcalfe2020}
{Metcalfe} T.~S.,  et~al., 2020, \mn@doi [\apj] {10.3847/1538-4357/aba963},
  \href {https://ui.adsabs.harvard.edu/abs/2020ApJ...900..154M} {900, 154}

\bibitem[\protect\citeauthoryear{Miller Bertolami}{2016}]{Miller2016}
  Miller Bertolami M.~M., 2016, A\&A, 588,
  A25. doi:10.1051/0004-6361/201526577
  
\bibitem[\protect\citeauthoryear{{Mittag}, {Schmitt}  \&
  {Schr{\"o}der}}{{Mittag} et~al.}{2013}]{Mittag2013}
{Mittag} M.,  {Schmitt} J.~H.~M.~M.,   {Schr{\"o}der} K.~P.,  2013, \mn@doi
  [\aap] {10.1051/0004-6361/201219868}, \href
  {https://ui.adsabs.harvard.edu/abs/2013A&A...549A.117M} {549, A117}

\bibitem[\protect\citeauthoryear{{Mohanty}, {Basri}, {Shu}, {Allard}  \&
  {Chabrier}}{{Mohanty} et~al.}{2002}]{mohantyetal02-1}
{Mohanty} S.,  {Basri} G.,  {Shu} F.,  {Allard} F.,   {Chabrier} G.,  2002,
  \mn@doi [\apj] {10.1086/339911}, \href
  {http://adsabs.harvard.edu/abs/2002ApJ...571..469M} {571, 469}

\bibitem[\protect\citeauthoryear{{Morales} et~al.,}{{Morales}
  et~al.}{2009}]{Morales2009}
{Morales} J.~C.,  et~al., 2009, \mn@doi [\apj] {10.1088/0004-637X/691/2/1400},
  \href {https://ui.adsabs.harvard.edu/abs/2009ApJ...691.1400M} {691, 1400}

\bibitem[\protect\citeauthoryear{{Morgan}, {West}, {Garc{\'e}s}, {Catal{\'a}n},
  {Dhital}, {Fuchs}  \& {Silvestri}}{{Morgan} et~al.}{2012}]{morganetal12-1}
{Morgan} D.~P.,  {West} A.~A.,  {Garc{\'e}s} A.,  {Catal{\'a}n} S.,  {Dhital}
  S.,  {Fuchs} M.,   {Silvestri} N.~M.,  2012, \mn@doi [\aj]
  {10.1088/0004-6256/144/4/93}, \href
  {http://adsabs.harvard.edu/abs/2012AJ....144...93M} {144, 93}

\bibitem[\protect\citeauthoryear{{Morin} et~al.,}{{Morin}
  et~al.}{2008}]{morinetal08-1}
{Morin} J.,  et~al., 2008, \mn@doi [\mnras] {10.1111/j.1365-2966.2008.13809.x},
  \href {http://adsabs.harvard.edu/abs/2008MNRAS.390..567M} {390, 567}

\bibitem[\protect\citeauthoryear{{Morin}, {Donati}, {Petit}, {Delfosse},
  {Forveille}  \& {Jardine}}{{Morin} et~al.}{2010}]{morinetal10-1}
{Morin} J.,  {Donati} J.-F.,  {Petit} P.,  {Delfosse} X.,  {Forveille} T.,
  {Jardine} M.~M.,  2010, \mn@doi [\mnras] {10.1111/j.1365-2966.2010.17101.x},
  \href {http://adsabs.harvard.edu/abs/2010MNRAS.407.2269M} {407, 2269}

\bibitem[\protect\citeauthoryear{{Moss} et~al.,}{{Moss}
  et~al.}{2022}]{Moss2022}
{Moss} A.,  et~al., 2022, \mn@doi [\apj] {10.3847/1538-4357/ac5ac0}, \href
  {https://ui.adsabs.harvard.edu/abs/2022ApJ...929...26M} {929, 26}

\bibitem[\protect\citeauthoryear{{Newton}, {Irwin}, {Charbonneau},
  {Berta-Thompson}, {Dittmann}  \& {West}}{{Newton} et~al.}{2016}]{Newton2016}
{Newton} E.~R.,  {Irwin} J.,  {Charbonneau} D.,  {Berta-Thompson} Z.~K.,
  {Dittmann} J.~A.,   {West} A.~A.,  2016, \mn@doi [\apj]
  {10.3847/0004-637X/821/2/93}, \href
  {https://ui.adsabs.harvard.edu/abs/2016ApJ...821...93N} {821, 93}

\bibitem[\protect\citeauthoryear{{Noyes}, {Hartmann}, {Baliunas}, {Duncan}  \&
  {Vaughan}}{{Noyes} et~al.}{1984}]{noyesetal84-1}
{Noyes} R.~W.,  {Hartmann} L.~W.,  {Baliunas} S.~L.,  {Duncan} D.~K.,
  {Vaughan} A.~H.,  1984, \mn@doi [\apj] {10.1086/161945}, \href
  {http://adsabs.harvard.edu/abs/1984ApJ...279..763N} {279, 763}

\bibitem[\protect\citeauthoryear{{Pace}}{{Pace}}{2013}]{Pace2013}
{Pace} G.,  2013, \mn@doi [\aap] {10.1051/0004-6361/201220364}, \href
  {https://ui.adsabs.harvard.edu/abs/2013A&A...551L...8P} {551, L8}

\bibitem[\protect\citeauthoryear{{Parker}}{{Parker}}{1955}]{parker55-1}
{Parker} E.~N.,  1955, \mn@doi [\apj] {10.1086/146087}, \href
  {http://adsabs.harvard.edu/abs/1955ApJ...122..293P} {122, 293}

\bibitem[\protect\citeauthoryear{{Pecaut} \& {Mamajek}}{{Pecaut} \&
  {Mamajek}}{2013}]{Pecaut2013}
{Pecaut} M.~J.,  {Mamajek} E.~E.,  2013, \mn@doi [\apjs]
  {10.1088/0067-0049/208/1/9}, \href
  {https://ui.adsabs.harvard.edu/abs/2013ApJS..208....9P} {208, 9}

\bibitem[\protect\citeauthoryear{Pietrinferni et
    al.}{2004}]{Pietrinferni2004} Pietrinferni A., Cassisi S., Salaris
  M., Castelli F., 2004, ApJ, 612, 168. doi:10.1086/422498
  
\bibitem[\protect\citeauthoryear{{Pineda}, {Youngblood}  \& {France}}{{Pineda}
  et~al.}{2021}]{Pineda2021}
{Pineda} J.~S.,  {Youngblood} A.,   {France} K.,  2021, arXiv e-prints, \href
  {https://ui.adsabs.harvard.edu/abs/2021arXiv210212485P} {p. arXiv:2102.12485}

\bibitem[\protect\citeauthoryear{{Pipin} \& {Yokoi}}{{Pipin} \&
  {Yokoi}}{2018}]{Pipin2018}
{Pipin} V.~V.,  {Yokoi} N.,  2018, \mn@doi [\apj] {10.3847/1538-4357/aabae6},
  \href {https://ui.adsabs.harvard.edu/abs/2018ApJ...859...18P} {859, 18}

\bibitem[\protect\citeauthoryear{{Pizzolato}, {Maggio}, {Micela}, {Sciortino}
  \& {Ventura}}{{Pizzolato} et~al.}{2003}]{pizzolatoetal03-1}
{Pizzolato} N.,  {Maggio} A.,  {Micela} G.,  {Sciortino} S.,   {Ventura} P.,
  2003, \mn@doi [] {10.1051/0004-6361:20021560}, \href {2003A&A...397..147P}
  {397, 147}

\bibitem[\protect\citeauthoryear{{Qiu}, {Tian}, {Wang}, {Nie}, {von Hippe},
  {Liu}, {Fouesneau}  \& {Rix}}{{Qiu} et~al.}{2020}]{Qiu2020}
{Qiu} D.,  {Tian} H.-J.,  {Wang} X.-D.,  {Nie} J.-L.,  {von Hippe} T.,  {Liu}
  G.-C.,  {Fouesneau} M.,   {Rix} H.-W.,  2020, arXiv e-prints, \href
  {https://ui.adsabs.harvard.edu/abs/2020arXiv201204890Q} {p. arXiv:2012.04890}

\bibitem[\protect\citeauthoryear{{Qiu}, {Tian}, {Wang}, {Nie}, {von Hippel},
  {Liu}, {Fouesneau}  \& {Rix}}{{Qiu} et~al.}{2021}]{Qiu2021}
{Qiu} D.,  {Tian} H.-J.,  {Wang} X.-D.,  {Nie} J.-L.,  {von Hippel} T.,  {Liu}
  G.-C.,  {Fouesneau} M.,   {Rix} H.-W.,  2021, \mn@doi [\apjs]
  {10.3847/1538-4365/abe468}, \href
  {https://ui.adsabs.harvard.edu/abs/2021ApJS..253...58Q} {253, 58}

\bibitem[\protect\citeauthoryear{{Raddi} et~al.,}{{Raddi}
  et~al.}{2022}]{Raddi2022}
{Raddi} R.,  et~al., 2022, \mn@doi [\aap] {10.1051/0004-6361/202141837}, \href
  {https://ui.adsabs.harvard.edu/abs/2022A&A...658A..22R} {658, A22}

\bibitem[\protect\citeauthoryear{{Raedler}, {Wiedemann}, {Brandenburg},
  {Meinel}  \& {Tuominen}}{{Raedler} et~al.}{1990}]{raedleretal90-1}
{Raedler} K.-H.,  {Wiedemann} E.,  {Brandenburg} A.,  {Meinel} R.,   {Tuominen}
  I.,  1990, \aap, \href {http://adsabs.harvard.edu/abs/1990A\%26A...239..413R}
  {239, 413}

\bibitem[\protect\citeauthoryear{{Rebassa-Mansergas}, {G{\"a}nsicke},
  {Rodr{\'\i}guez-Gil}, {Schreiber}  \& {Koester}}{{Rebassa-Mansergas}
  et~al.}{2007}]{Rebassa2007}
{Rebassa-Mansergas} A.,  {G{\"a}nsicke} B.~T.,  {Rodr{\'\i}guez-Gil} P.,
  {Schreiber} M.~R.,   {Koester} D.,  2007, \mn@doi [\mnras]
  {10.1111/j.1365-2966.2007.12288.x}, \href
  {https://ui.adsabs.harvard.edu/abs/2007MNRAS.382.1377R} {382, 1377}

\bibitem[\protect\citeauthoryear{{Rebassa-Mansergas}, {G{\"a}nsicke},
  {Schreiber}, {Koester}  \& {Rodr{\'{\i}}guez-Gil}}{{Rebassa-Mansergas}
  et~al.}{2010}]{rebassa-mansergasetal10-1}
{Rebassa-Mansergas} A.,  {G{\"a}nsicke} B.~T.,  {Schreiber} M.~R.,  {Koester}
  D.,   {Rodr{\'{\i}}guez-Gil} P.,  2010, \mn@doi [\mnras]
  {10.1111/j.1365-2966.2009.15915.x}, \href
  {http://adsabs.harvard.edu/abs/2010MNRAS.402..620R} {402, 620}

\bibitem[\protect\citeauthoryear{{Rebassa-Mansergas}, {Nebot
  G{\'o}mez-Mor{\'a}n}, {Schreiber}, {G{\"a}nsicke}, {Schwope}, {Gallardo}  \&
  {Koester}}{{Rebassa-Mansergas} et~al.}{2012}]{rebassa-mansergasetal12-1}
{Rebassa-Mansergas} A.,  {Nebot G{\'o}mez-Mor{\'a}n} A.,  {Schreiber} M.~R.,
  {G{\"a}nsicke} B.~T.,  {Schwope} A.,  {Gallardo} J.,   {Koester} D.,  2012,
  \mn@doi [\mnras] {10.1111/j.1365-2966.2011.19923.x}, \href
  {http://adsabs.harvard.edu/abs/2012MNRAS.419..806R} {419, 806}

\bibitem[\protect\citeauthoryear{{Rebassa-Mansergas}, {Schreiber}  \&
  {G{\"a}nsicke}}{{Rebassa-Mansergas}
  et~al.}{2013a}]{rebassa-mansergasetal13-1}
{Rebassa-Mansergas} A.,  {Schreiber} M.~R.,   {G{\"a}nsicke} B.~T.,  2013a,
  \mn@doi [\mnras] {10.1093/mnras/sts630}, \href
  {http://adsabs.harvard.edu/abs/2013MNRAS.429.3570R} {429, 3570}

\bibitem[\protect\citeauthoryear{{Rebassa-Mansergas}, {Agurto-Gangas},
  {Schreiber}, {G{\"a}nsicke}  \& {Koester}}{{Rebassa-Mansergas}
  et~al.}{2013b}]{rebassa-mansergasetal13-2}
{Rebassa-Mansergas} A.,  {Agurto-Gangas} C.,  {Schreiber} M.~R.,
  {G{\"a}nsicke} B.~T.,   {Koester} D.,  2013b, \mn@doi [\mnras]
  {10.1093/mnras/stt974}, \href
  {http://adsabs.harvard.edu/abs/2013MNRAS.433.3398R} {433, 3398}

\bibitem[\protect\citeauthoryear{{Rebassa-Mansergas}, {Ren}, {Parsons},
  {G{\"a}nsicke}, {Schreiber}, {Garc{\'{\i}}a-Berro}, {Liu}  \&
  {Koester}}{{Rebassa-Mansergas} et~al.}{2016a}]{rebassa-mansergasetal16-1}
{Rebassa-Mansergas} A.,  {Ren} J.~J.,  {Parsons} S.~G.,  {G{\"a}nsicke} B.~T.,
  {Schreiber} M.~R.,  {Garc{\'{\i}}a-Berro} E.,  {Liu} X.-W.,   {Koester} D.,
  2016a, \mn@doi [\mnras] {10.1093/mnras/stw554}, \href
  {http://adsabs.harvard.edu/abs/2016MNRAS.458.3808R} {458, 3808}

\bibitem[\protect\citeauthoryear{{Rebassa-Mansergas}
  et~al.,}{{Rebassa-Mansergas} et~al.}{2016b}]{Rebassa2016b}
{Rebassa-Mansergas} A.,  et~al., 2016b, \mn@doi [\mnras]
  {10.1093/mnras/stw2021}, \href
  {https://ui.adsabs.harvard.edu/abs/2016MNRAS.463.1137R} {463, 1137}

\bibitem[\protect\citeauthoryear{{Rebassa-Mansergas}
  et~al.,}{{Rebassa-Mansergas} et~al.}{2021}]{rebassa-mansergasetal21}
{Rebassa-Mansergas} A.,  et~al., 2021, \mn@doi [\mnras]
  {10.1093/mnras/stab1559}, \href
  {https://ui.adsabs.harvard.edu/abs/2021MNRAS.505.3165R} {505, 3165}

\bibitem[\protect\citeauthoryear{{Reiners} \& {Basri}}{{Reiners} \&
  {Basri}}{2008}]{reiners+basri08-1}
{Reiners} A.,  {Basri} G.,  2008, \mn@doi [\apj] {10.1086/590073}, \href
  {http://adsabs.harvard.edu/abs/2008ApJ...684.1390R} {684, 1390}

\bibitem[\protect\citeauthoryear{{Reiners} \& {Basri}}{{Reiners} \&
  {Basri}}{2009}]{reiners+basri09-1}
{Reiners} A.,  {Basri} G.,  2009, \mn@doi [\aap] {10.1051/0004-6361:200811450},
  \href {http://adsabs.harvard.edu/abs/2009A\%26A...496..787R} {496, 787}

\bibitem[\protect\citeauthoryear{{Reiners} \& {Basri}}{{Reiners} \&
  {Basri}}{2010}]{reiners+basri10-1}
{Reiners} A.,  {Basri} G.,  2010, \mn@doi [\apj] {10.1088/0004-637X/710/2/924},
  \href {http://adsabs.harvard.edu/abs/2010ApJ...710..924R} {710, 924}

\bibitem[\protect\citeauthoryear{{Reiners} \& {Mohanty}}{{Reiners} \&
  {Mohanty}}{2012}]{reiners+mohanty12-1}
{Reiners} A.,  {Mohanty} S.,  2012, \mn@doi [\apj]
  {10.1088/0004-637X/746/1/43}, \href
  {http://adsabs.harvard.edu/abs/2012ApJ...746...43R} {746, 43}

\bibitem[\protect\citeauthoryear{{Reiners}, {Basri}  \& {Browning}}{{Reiners}
  et~al.}{2009}]{reinersetal09-1}
{Reiners} A.,  {Basri} G.,   {Browning} M.,  2009, \mn@doi [\apj]
  {10.1088/0004-637X/692/1/538}, \href
  {http://adsabs.harvard.edu/abs/2009ApJ...692..538R} {692, 538}

\bibitem[\protect\citeauthoryear{{Reiners}, {Joshi}  \& {Goldman}}{{Reiners}
  et~al.}{2012}]{Reiners2012}
{Reiners} A.,  {Joshi} N.,   {Goldman} B.,  2012, \mn@doi [\aj]
  {10.1088/0004-6256/143/4/93}, \href
  {https://ui.adsabs.harvard.edu/abs/2012AJ....143...93R} {143, 93}

\bibitem[\protect\citeauthoryear{{Renedo}, {Althaus}, {Miller Bertolami},
  {Romero}, {C{\'o}rsico}, {Rohrmann}  \& {Garc{\'\i}a-Berro}}{{Renedo}
  et~al.}{2010}]{Renedo2010}
{Renedo} I.,  {Althaus} L.~G.,  {Miller Bertolami} M.~M.,  {Romero} A.~D.,
  {C{\'o}rsico} A.~H.,  {Rohrmann} R.~D.,   {Garc{\'\i}a-Berro} E.,  2010,
  \mn@doi [\apj] {10.1088/0004-637X/717/1/183}, \href
  {https://ui.adsabs.harvard.edu/abs/2010ApJ...717..183R} {717, 183}

\bibitem[\protect\citeauthoryear{{R{\'e}ville}, {Brun}, {Matt}, {Strugarek}  \&
  {Pinto}}{{R{\'e}ville} et~al.}{2015}]{Reville2015}
{R{\'e}ville} V.,  {Brun} A.~S.,  {Matt} S.~P.,  {Strugarek} A.,   {Pinto}
  R.~F.,  2015, \mn@doi [\apj] {10.1088/0004-637X/798/2/116}, \href
  {https://ui.adsabs.harvard.edu/abs/2015ApJ...798..116R} {798, 116}

\bibitem[\protect\citeauthoryear{{Sadeghi Ardestani}, {Guillot}  \&
  {Morel}}{{Sadeghi Ardestani} et~al.}{2017}]{Sadeghi2017}
{Sadeghi Ardestani} L.,  {Guillot} T.,   {Morel} P.,  2017, \mn@doi [\mnras]
  {10.1093/mnras/stx2039}, \href
  {https://ui.adsabs.harvard.edu/abs/2017MNRAS.472.2590S} {472, 2590}

\bibitem[\protect\citeauthoryear{{Schreiber} et~al.,}{{Schreiber}
  et~al.}{2010}]{Schreiber2010}
{Schreiber} M.~R.,  et~al., 2010, \mn@doi [\aap] {10.1051/0004-6361/201013990},
  \href {https://ui.adsabs.harvard.edu/abs/2010A&A...513L...7S} {513, L7}

\bibitem[\protect\citeauthoryear{{Shulyak}, {Reiners}, {Engeln}, {Malo},
  {Yadav}, {Morin}  \& {Kochukhov}}{{Shulyak} et~al.}{2017}]{Shulyak2017}
{Shulyak} D.,  {Reiners} A.,  {Engeln} A.,  {Malo} L.,  {Yadav} R.,  {Morin}
  J.,   {Kochukhov} O.,  2017, \mn@doi [Nature Astronomy]
  {10.1038/s41550-017-0184}, \href
  {https://ui.adsabs.harvard.edu/abs/2017NatAs...1E.184S} {1, 0184}

\bibitem[\protect\citeauthoryear{{Sills}, {Pinsonneault}  \&
  {Terndrup}}{{Sills} et~al.}{2000}]{sillsetal00-1}
{Sills} A.,  {Pinsonneault} M.~H.,   {Terndrup} D.~M.,  2000, ApJ, 534, 335

\bibitem[\protect\citeauthoryear{{Silva Aguirre} et~al.,}{{Silva Aguirre}
  et~al.}{2017}]{Silva2017}
{Silva Aguirre} V.,  et~al., 2017, \mn@doi [\apj]
  {10.3847/1538-4357/835/2/173}, \href
  {https://ui.adsabs.harvard.edu/abs/2017ApJ...835..173S} {835, 173}

\bibitem[\protect\citeauthoryear{{Skinner}, {Morgan}, {West}, {L{\'e}pine}  \&
  {Thorstensen}}{{Skinner} et~al.}{2017}]{Skinner2017}
{Skinner} J.~N.,  {Morgan} D.~P.,  {West} A.~A.,  {L{\'e}pine} S.,
  {Thorstensen} J.~R.,  2017, \mn@doi [\aj] {10.3847/1538-3881/aa83b5}, \href
  {https://ui.adsabs.harvard.edu/abs/2017AJ....154..118S} {154, 118}

\bibitem[\protect\citeauthoryear{{Skumanich}}{{Skumanich}}{1972}]{skumanich72-1}
{Skumanich} A.,  1972, \mn@doi [\apj] {10.1086/151310}, \href
  {http://adsabs.harvard.edu/abs/1972ApJ...171..565S} {171, 565}

\bibitem[\protect\citeauthoryear{{Soderblom}}{{Soderblom}}{2010}]{soderblom10-1}
{Soderblom} D.~R.,  2010, \mn@doi [\araa]
  {10.1146/annurev-astro-081309-130806}, \href
  {http://adsabs.harvard.edu/abs/2010ARA\%26A..48..581S} {48, 581}

\bibitem[\protect\citeauthoryear{{Spiegel} \& {Zahn}}{{Spiegel} \&
  {Zahn}}{1992}]{spiegel+zahn92-1}
{Spiegel} E.~A.,  {Zahn} J.-P.,  1992, \aap, 265, 106

\bibitem[\protect\citeauthoryear{{Su{\'a}rez Mascare{\~n}o}, {Rebolo},
  {Gonz{\'a}lez Hern{\'a}ndez}  \& {Esposito}}{{Su{\'a}rez Mascare{\~n}o}
  et~al.}{2015}]{Suarez2015}
{Su{\'a}rez Mascare{\~n}o} A.,  {Rebolo} R.,  {Gonz{\'a}lez Hern{\'a}ndez}
  J.~I.,   {Esposito} M.,  2015, \mn@doi [\mnras] {10.1093/mnras/stv1441},
  \href {https://ui.adsabs.harvard.edu/abs/2015MNRAS.452.2745S} {452, 2745}

\bibitem[\protect\citeauthoryear{{Toloza} et~al.,}{{Toloza}
  et~al.}{2023}]{WDB2023}
{Toloza} O.,  et~al., 2023, \mn@doi [The Messenger] {10.18727/0722-6691/5299},
  \href {https://ui.adsabs.harvard.edu/abs/2023Msngr.190....4T} {190, 4}

\bibitem[\protect\citeauthoryear{{Torres}, {Garc{\'\i}a-Berro}, {Althaus}  \&
  {Camisassa}}{{Torres} et~al.}{2015}]{Torres2015}
{Torres} S.,  {Garc{\'\i}a-Berro} E.,  {Althaus} L.~G.,   {Camisassa} M.~E.,
  2015, \mn@doi [\aap] {10.1051/0004-6361/201526157}, \href
  {https://ui.adsabs.harvard.edu/abs/2015A&A...581A..90T} {581, A90}

\bibitem[\protect\citeauthoryear{{Torres} et~al.,}{{Torres}
  et~al.}{2023}]{Torres2023}
{Torres} S.,  et~al., 2023, \mn@doi [arXiv e-prints]
  {10.48550/arXiv.2307.13629}, \href
  {https://ui.adsabs.harvard.edu/abs/2023arXiv230713629T} {p. arXiv:2307.13629}

\bibitem[\protect\citeauthoryear{{Tuffs}, {Popescu}, {V{\"o}lk}, {Kylafis}  \&
  {Dopita}}{{Tuffs} et~al.}{2004}]{Tuffs2004}
{Tuffs} R.~J.,  {Popescu} C.~C.,  {V{\"o}lk} H.~J.,  {Kylafis} N.~D.,
  {Dopita} M.~A.,  2004, \mn@doi [\aap] {10.1051/0004-6361:20035689}, \href
  {https://ui.adsabs.harvard.edu/abs/2004A&A...419..821T} {419, 821}

\bibitem[\protect\citeauthoryear{{Vaughan}, {Preston}  \& {Wilson}}{{Vaughan}
  et~al.}{1978}]{Vaughan1978}
{Vaughan} A.~H.,  {Preston} G.~W.,   {Wilson} O.~C.,  1978, \mn@doi [\pasp]
  {10.1086/130324}, \href
  {https://ui.adsabs.harvard.edu/abs/1978PASP...90..267V} {90, 267}

\bibitem[\protect\citeauthoryear{{West} \& {Basri}}{{West} \&
  {Basri}}{2009}]{west+basri09-1}
{West} A.~A.,  {Basri} G.,  2009, \mn@doi [\apj]
  {10.1088/0004-637X/693/2/1283}, \href
  {http://adsabs.harvard.edu/abs/2009ApJ...693.1283W} {693, 1283}

\bibitem[\protect\citeauthoryear{{West}, {Hawley}, {Bochanski}, {Covey},
  {Reid}, {Dhital}, {Hilton}  \& {Masuda}}{{West} et~al.}{2008}]{westetal08-1}
{West} A.~A.,  {Hawley} S.~L.,  {Bochanski} J.~J.,  {Covey} K.~R.,  {Reid}
  I.~N.,  {Dhital} S.,  {Hilton} E.~J.,   {Masuda} M.,  2008, \mn@doi [\aj]
  {10.1088/0004-6256/135/3/785}, \href
  {http://adsabs.harvard.edu/abs/2008AJ....135..785W} {135, 785}

\bibitem[\protect\citeauthoryear{{Wilson}}{{Wilson}}{1966}]{wilson66-1}
{Wilson} O.~C.,  1966, \mn@doi [\apj] {10.1086/148649}, \href
  {http://adsabs.harvard.edu/abs/1966ApJ...144..695W} {144, 695}

\bibitem[\protect\citeauthoryear{{Wilson}}{{Wilson}}{1968}]{Wilson1968}
{Wilson} O.~C.,  1968, \mn@doi [\apj] {10.1086/149652}, \href
  {https://ui.adsabs.harvard.edu/abs/1968ApJ...153..221W} {153, 221}

\bibitem[\protect\citeauthoryear{{Wilson}}{{Wilson}}{1978}]{Wilson1978}
{Wilson} O.~C.,  1978, \mn@doi [\apj] {10.1086/156618}, \href
  {https://ui.adsabs.harvard.edu/abs/1978ApJ...226..379W} {226, 379}

\bibitem[\protect\citeauthoryear{{Wright}, {Newton}, {Williams}, {Drake}  \&
  {Yadav}}{{Wright} et~al.}{2018}]{Wright2018}
{Wright} N.~J.,  {Newton} E.~R.,  {Williams} P. K.~G.,  {Drake} J.~J.,
  {Yadav} R.~K.,  2018, \mn@doi [\mnras] {10.1093/mnras/sty1670}, \href
  {https://ui.adsabs.harvard.edu/abs/2018MNRAS.479.2351W} {479, 2351}

\bibitem[\protect\citeauthoryear{{Yuxi}, {Lu}, {Angus}, {Curtis}, {David}  \&
  {Kiman}}{{Yuxi} et~al.}{2021}]{Yuxi2021}
{Yuxi} {Lu} {Angus} R.,  {Curtis} J.~L.,  {David} T.~J.,   {Kiman} R.,  2021,
  arXiv e-prints, \href {https://ui.adsabs.harvard.edu/abs/2021arXiv210201772Y}
  {p. arXiv:2102.01772}

\bibitem[\protect\citeauthoryear{{Zhao}, {Oswalt}, {Rudkin}, {Zhao}  \&
  {Chen}}{{Zhao} et~al.}{2011}]{Zhao2011}
{Zhao} J.~K.,  {Oswalt} T.~D.,  {Rudkin} M.,  {Zhao} G.,   {Chen} Y.~Q.,  2011,
  \mn@doi [\aj] {10.1088/0004-6256/141/4/107}, \href
  {https://ui.adsabs.harvard.edu/abs/2011AJ....141..107Z} {141, 107}

\bibitem[\protect\citeauthoryear{{Zhao} et~al.,}{{Zhao}
  et~al.}{2015}]{Zhao2015}
{Zhao} J.-K.,  et~al., 2015, \mn@doi [Research in Astronomy and Astrophysics]
  {10.1088/1674-4527/15/8/013}, \href
  {https://ui.adsabs.harvard.edu/abs/2015RAA....15.1282Z} {15, 1282}

\bibitem[\protect\citeauthoryear{{Zorotovic} et~al.,}{{Zorotovic}
  et~al.}{2016}]{Zorotovic2016}
{Zorotovic} M.,  et~al., 2016, \mn@doi [\mnras] {10.1093/mnras/stw246}, \href
  {https://ui.adsabs.harvard.edu/abs/2016MNRAS.457.3867Z} {457, 3867}

\bibitem[\protect\citeauthoryear{{Zurlo} et~al.,}{{Zurlo}
  et~al.}{2013}]{Zurlo2013}
{Zurlo} A.,  et~al., 2013, \mn@doi [\aap] {10.1051/0004-6361/201321179}, \href
  {https://ui.adsabs.harvard.edu/abs/2013A&A...554A..21Z} {554, A21}

\bibitem[\protect\citeauthoryear{{de Jong} et~al.,}{{de Jong}
  et~al.}{2022}]{4MOST}
{de Jong} R.~S.,  et~al., 2022, in {Evans} C.~J.,  {Bryant} J.~J.,   {Motohara}
  K.,  eds,  Society of Photo-Optical Instrumentation Engineers (SPIE)
  Conference Series Vol. 12184, Society of Photo-Optical Instrumentation
  Engineers (SPIE) Conference Series. p. 1218414, \mn@doi{10.1117/12.2628965}

\bibitem[\protect\citeauthoryear{{van Saders}, {Ceillier}, {Metcalfe}, {Silva
  Aguirre}, {Pinsonneault}, {Garc{\'\i}a}, {Mathur}  \& {Davies}}{{van Saders}
  et~al.}{2016}]{vanSaders2016}
{van Saders} J.~L.,  {Ceillier} T.,  {Metcalfe} T.~S.,  {Silva Aguirre} V.,
  {Pinsonneault} M.~H.,  {Garc{\'\i}a} R.~A.,  {Mathur} S.,   {Davies} G.~R.,
  2016, \mn@doi [\nat] {10.1038/nature16168}, \href
  {https://ui.adsabs.harvard.edu/abs/2016Natur.529..181V} {529, 181}

\makeatother
\end{thebibliography}




\appendix

\section{Converting the S$_\mathrm{HK}$ index to the Mount Wilson scale}

\begin{figure}
    \centering
    \includegraphics[angle=-90,width=\columnwidth]{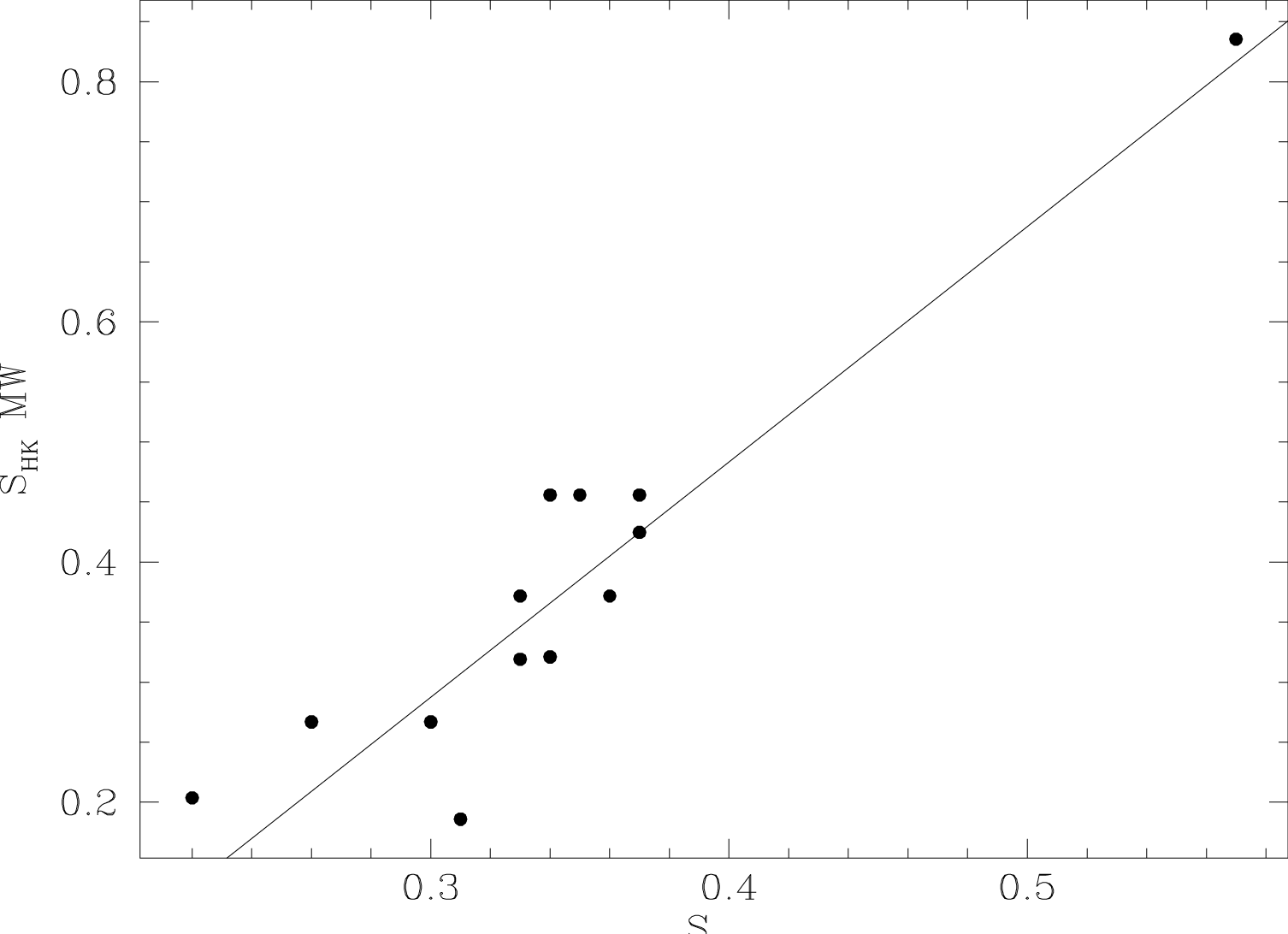}
    \caption{The S$_\mathrm{HK}$ index in the  Mount Wilson scale as a
      function of  the S$_\mathrm{HK}$  index directly measured  by us
      for  nine  stars  which   have  common  TNG/HARPS-N  and  LAMOST
      low-resolution spectra.}
    \label{fig:MW}
\end{figure}

We obtained the S$_\mathrm{HK}$ index in  the Mount Wilson scale of 27
stars  that were  observed with  the HARPS-N  spectrograph of  the TNG
using    the   routine    developed   by    \citet{Maldonado22}   (see
Section\,\ref{s-actrot}). Nine  of these  stars were also  observed by
LAMOST  with  the   low-resolution  instrument.   The  S$_\mathrm{HK}$
indexes      we      obtained      for      these      nine      stars
(Section\,\ref{sec:indicators})  are  compared  to  the  corresponding
values in  the Mount  Wilson scale  in Figure\,\ref{fig:MW}.   A least
squares   fit   to   the   data  results   in   S$_\mathrm{HK,MW}$   =
(1.52$\pm$0.10) $\times$ S$_\mathrm{HK}$ - (0.074$\pm$0.025).  We used
this equation to obtain the  S$_\mathrm{HK}$ in the Mount Wilson scale
of  123 additional  stars observe  by LAMOST  with its  low-resolution
spectrograph.

\bsp	
\label{lastpage}
\end{document}